# Exact Spectral Analysis of Single-$h$ and Multi-$h$ CPM Signals through PAM decomposition and Matrix Sereies Evaluation

G. Cariolaro, *Member IEEE*, T. Erseghe, N. Laurenti, G. Pierobon

# I. INTRODUCTION

Continuous phase modulation (CPM) is a widely used form of digital modulation which is employed in many modern standards due to its very good power and spectral efficiency [1]. CPM is currently used in 2nd generation GSM (global system for mobile communications) cell phones, but also in IEEE 802.11 FHSS, Bluetooth, and other proprietary wireless modems.

The (single-$h$) CPM signal is defined by

$$v(t) = e^{j\alpha(t)} \tag{1}$$

where

$$\alpha(t) = 2\pi\, h \sum_n a_n\, \varphi(t - nT) \tag{1a}$$

with: $h$ the *modulation index*; $\{a_n\}$ the data sequence with period $T$ and $M$–ary alphabet $\mathcal{A}_M = \{\pm 1, \pm 3, \ldots, \pm(M-1)\}$; $\varphi(t)$ the *phase response*, a monotonic function satisfying the property $\varphi(t) = 0$ for $t \le 0$ and $\varphi(t) = \frac{1}{2}$ for $t \ge LT$, with $L$ the system memory.

Thanks to its constant-envelope waveform, CPM guarantees a constant transmitted power, thus allowing use of low–cost nonlinear amplifiers, with a save of 3–10 dB loss of nonlinear distortion [1]. Moreover, due to the phase continuity, the CPM signals have good spectral properties compared with memoryless modulations. The attractive feautures of CPM must face some major drawbacks, the most important being the high implementation complexity required for an optimal receiver, and the even more challenging task of receiver optimization in nonlinear and/or fading channels [2], such as satellite and mobile radio channels. A common approach to complexity reduction, proposed by Laurent in 1986 [3], is that of interpreting CPM as the cascade of an encoder followed by a bank of PAM (pulse amplitude modulation) interpolators, some of which carry very minor portions of the signal power and can thus be discarded. In this way, moderate–complexity efficient receiver can be built. Many different approaches of PAM decomposition have been proposed during years [4], [5], [6], [7], making CPM still an attractive research issue.

More recently, the literature has been focusing on a generalization of the ordinary single–$h$ CPM format (1), formerly introduced by Anderson and Taylor in 1978 [8]. This generalization, known as multi–$h$ CPM, assumes that the modulation index is time–varying, to have

$$\alpha(t) = 2\pi \sum_n a_n\, h_n\, \varphi(t - nT) \tag{2}$$

where $\{h_n\}$ is usually a periodic sequence of period $N_h$. In comparison with single–$h$ CPM, the multi–$h$ CPM signal is known to be able to reduce the bandwidth [10] and increase the probability of detection at the receiver side [1], [11]. Also, low–complexity receivers through PAM decomposition of multi–$h$ CPM signals have already been proposed in [7], [12],

In this paper, we deal with another interesting aspect of multi–$h$ CPM signals, that is power spectral density (PSD) evaluation, an issue which has been investigated both in the past (e.g. see [9] and references therein) and in recent times [11], [13]. However, unlike the existing literature that has focused on numerical calculation, we aim at *closed form* PSD evaluation.

This is possible by interpreting the CPM signal as a combination of PAM signals, using Laurent approach. The key point is then to model the encoder, preceeding the bank of PAM interpolators, as a finite-state sequential machine (SM) [14] whose output PSD can be derived in closed–form by exploiting the results of [15]. As will be clearer later on, the direct application of [15] is not possible in the multi–$h$ CPM context, since that theory is valid for a *stationary* and *irreducible* SM, whereas multi–$h$ CPM has a *periodically time invariant* (PTI) structure. Furthermore, the peculiar input alphabet $\mathcal{A}_M$ leads to a *reducible* SM. In any case, a judicious modification of the SM case finally permits relying on [15]. Incidentally, the reducibility problem is also found with the ordinary (single-$h$) CPM case, of which we give new further insights. We also underline that the proposed method is closely related to the results of [16], with the non trivial differences of being able to propose a *truly closed form* and more compact results, and of being able to deal with a reducible SM.

The paper is organized as follows. Section II deals with the representation of a multi–$h$ CPM modulator through a finite–state SM, while the investigation of PTI and reducible structures is given in Section III. Then, Section IV illustrates the formulation of a *stationary* and *irreducible* SM generating the multi–$h$ CPM signal after a proper bank of PAM interpolators. Section V recalls the findings of [15], and uses them to formulate the closed–form PSD evaluation. Some examples of spectra are reported in Section VI. Finally, Section VII concludes the paper.

## II. Representation of CPM Modulator

In this section we develop a convenient representation of a multi–$h$ CPM modulator, consisting of a SM followed by a bank of interpolators. The deduction of the CPM representation is based on the basic decomposition (BD) of the CPM signal into PAM waveforms [6]. The periodically varying modulation index leads to a PTI SM and to PTI interpolators.

### A. The basic decomposition for multi–$h$ CPM

The phase of the multi–$h$ CPM signal (2) in the interval $\mathcal{I}_n = [nT, nT+T)$ can be written in the form

$$\alpha(t) = 2\pi \left[ \sum_{m=-\infty}^{n-L} a_m h_m \frac{1}{2} + \sum_{m=n-L+1}^{n} a_m h_m \, \varphi(t - mT) \right] . \tag{3}$$

Then

$$v(t) = \sigma_{n-L} \prod_{i=0}^{L-1} e^{j a_{n-i} h_{n-i} \varphi(t-(n-i)T)} , \qquad t \in \mathcal{I}_n \tag{4}$$

where

$$\sigma_{n-L} = \prod_{m=-\infty}^{n-L} J_m^{a_m} \quad \text{with} \quad J_m = e^{j\pi h_m} \tag{5}$$

plays the role of a *state*, which is renewed according to the relation

$$\sigma_{m+1} = \sigma_m J_{m+1}^{a_{m+1}} . \tag{6}$$

Comparison with the single–$h$ case [14] shows that the BD takes the form

$$v(t) = \sum_n \boldsymbol{q}_n(t - nT) \, \boldsymbol{b}_n \tag{7}$$

where

$$\boldsymbol{b}_n = \sigma_{n-L} \left[ \delta_{\boldsymbol{a}_n, \boldsymbol{\alpha}} \right]_{\boldsymbol{\alpha} \in \mathcal{A}_M^L} \tag{7a}$$

is a *word sequence* (column vectors of length $N = M^L$), and

$$\boldsymbol{q}_n(t) = \eta_T(t) \left[ \prod_{i=0}^{L-1} \exp\left( j 2\pi h_{n-i} \alpha_i \varphi(t + iT) \right) \right]_{\boldsymbol{\alpha} \in \mathcal{A}_M^L} \tag{7b}$$

constitutes an interpolating filter bank (row vectors of length $N$). Here, $\eta_T(t)$ is an indicator function active over the interval $[0, T)$, $\delta$ is a vector generalization of the Kronecker delta function, $\boldsymbol{a}_n = (a_{n-L+1}, \dots, a_n)$ collects the input data over a window of length $L$, $\boldsymbol{\alpha} = (\alpha_0, \alpha_1, \dots, \alpha_{L-1}) \in \mathcal{A}_M^L$, and $N$ is the cardinality of $\mathcal{A}_M^L$.

The difference with respect to the single–$h$ CPM is in the renewal law (5), which is periodically time–dependent (PTI), and in the pulses $\boldsymbol{q}_n(t)$, which are PTI with respect to the time $n$. This is shown in Fig. 1, where the non–linear device produces the symbols $\boldsymbol{b}_n$ from the input data and the bank of interpolators produces the PAM waveforms.

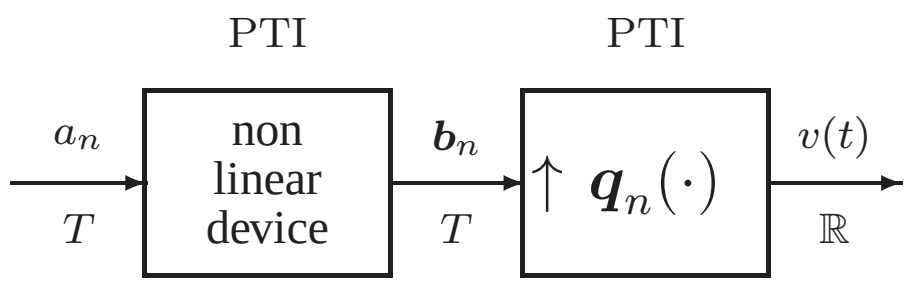

Fig. 1 – A first model of a multi–$h$ CPM modulator.

We now discuss the nature of the "state" $\sigma_m$, whose renewal law is given by (6). We suppose that *all the modulation indexes are rational* and we write them in the form $h_i = r_i/p$, where $p$ is the least common denominator of the fractions. As an example [7], if $h_0 = \frac{1}{4} = \frac{4}{16}$, $h_1 = \frac{5}{16}$, $h_2 = \frac{8}{16}$, $h_3 = \frac{5}{8} = \frac{10}{16}$, the integer $p$ is $16$. Then we can see that

$$\sigma_m \in \left\{1, W_{2p}, W_{2p}^2, \ldots, W_{2p}^{2p-1}\right\} \triangleq \mathcal{W}_{2p} \ , \ W_{2p} = e^{j2\pi/(2p)} \ . \tag{8}$$

It will be convenient to replace $\sigma_m$ by an equivalent state $z_m$ such that

$$\sigma_m = W_{2p}^{z_m} \tag{9}$$

which takes the values in the integer set

$$z_m \in \{0, 1, \ldots, 2p-1\} \triangleq \mathcal{N}_{2p} \ . \tag{10}$$

The renewal law for $z_m$ is obtained by rewriting (5) using (9), namely

$$W_{2p}^{s_{m+1}} = W_{2p}^{z_m} \, e^{j\pi r_{m+1} a_{m+1}/p} = W_{2p}^{z_m + r_{m+1} a_{m+1}} \ .$$

Hence

$$z_{m+1} = (z_m + r_{m+1} a_{m+1})_{2p} \tag{11}$$

where $(\ )_{2p}$ denotes modulo $2p$.

In terms of $z_m$ the word sequence becomes

$$\begin{aligned} \boldsymbol{b}_n &= W_{2p}^{z_{n-L}} \left[\delta_{\boldsymbol{a}_n, \boldsymbol{\alpha}}\right]_{\boldsymbol{\alpha} \in \mathcal{A}_M^L} \\ &= W_{2p}^{z_{n-L}} \left[\delta_{a_{n-L+1}, \alpha_0} \cdots \delta_{a_n, \alpha_{L-1}}\right]_{\boldsymbol{\alpha} \in \mathcal{A}_M^L} \ . \end{aligned} \tag{12}$$

Now, the vector $\boldsymbol{b}_n$ of length $N = M^L$ can be conveniently written by the Kronecker product[1] $\otimes$. Specifically,

PROPOSITION 1 Let $\boldsymbol{w}_{a_n} = [\delta_{a_n,\alpha}]_{\alpha\in\mathcal{A}_M}$ be the indicator vector of $a_n$ (a column vector of size $M$). Then

$$\boldsymbol{b}_n = W_{2p}^{z_{n-L}} \boldsymbol{w}_{a_{n-L+1}} \otimes \cdots \otimes \boldsymbol{w}_{a_{n-1}} \otimes \boldsymbol{w}_{a_n} \ . \qquad \square \tag{13}$$

For instance, for $L = 1$ and $M = 4$ we have

$$[\delta_{a_n,\alpha}]_{\alpha\in\mathcal{A}_4} = W_{2p}^{z_{n-1}} \boldsymbol{w}_{a_n}$$

where

$$\boldsymbol{w}_{-3} = [1 \quad 0 \quad 0 \quad 0]^T , \qquad \boldsymbol{w}_{-1} = [0 \quad 1 \quad 0 \quad 0]^T$$
$$\boldsymbol{w}_{+1} = [0 \quad 0 \quad 1 \quad 0]^T , \qquad \boldsymbol{w}_{+3} = [0 \quad 0 \quad 0 \quad 1]^T .$$

### *B. Formulation of the sequential machine*

We now formalize the non–linear device of Fig. 1 as a PTI SM (see [14] for the single–$h$ case), that is as a quintuple

$$\mathcal{M}_{\text{CPM}} = (\mathcal{A}, \mathcal{B}, \mathcal{S}, \boldsymbol{g}, \boldsymbol{h}) \tag{14}$$

where $\mathcal{A}$ is the input alphabet, $\mathcal{B}$ is the output alphabet, $\mathcal{S}$ is the state alphabet, $\boldsymbol{g}$ is the state–update function and $\boldsymbol{h}$ is the output function, with

$$\begin{cases} \boldsymbol{s}_{n+1} = \boldsymbol{g}\Big(\boldsymbol{s}_n, \boldsymbol{a}_n, n\Big) \ , & \boldsymbol{a}_n \in \mathcal{A}\,,\ \boldsymbol{s}_n, \boldsymbol{s}_{n+1} \in \mathcal{S} \\ \boldsymbol{b}_n = \boldsymbol{h}\Big(\boldsymbol{s}_n, \boldsymbol{a}_n, n\Big) \ , & \boldsymbol{b}_n \in \mathcal{B}\,. \end{cases} \tag{15}$$

Note that, in general, the state update function and the output function in (15) depend on $n$. When the dependence on $n$ can be dropped we will talk of a stationary update/output function, and when the dependence is periodic on $n$ we will talk of a PTI update/output function.

[1]Given two matrices $\mathbf{A} = ||a_{mn}||$ and $\mathbf{B} = ||b_{ij}||$ of arbitrary dimensions $M_A \times N_A$ and $M_B \times N_B$, the *Kronecker product* $\mathbf{A} \times \mathbf{B}$ is defined by

$$\mathbf{A} \times \mathbf{B} = \begin{bmatrix} a_{11}\,\mathbf{B} & \dots & a_{1N_A}\,\mathbf{B} \\ \vdots & & \vdots \\ a_{M_A1}\,\mathbf{B} & \dots & a_{M_AN_A}\,\mathbf{B} \end{bmatrix}$$

and is a matrix of dimensions $(M_A M_B) \times (N_A N_B)$. The $L$th *Kronecker power* of a matrix $\mathbf{A}$ is defined as $\mathbf{A}^{\times L} = \mathbf{A} \times \mathbf{A} \times \cdots \times \mathbf{A}$ ($L$ factors). The following property relates the Kronecker product to the ordinary matrix product (*mixed–product law*) $(\mathbf{A} \times \mathbf{B})\,(\mathbf{C} \times \mathbf{D}) = (\mathbf{A}\,\mathbf{C}) \times (\mathbf{B}\,\mathbf{D})$ and more generally $(\mathbf{A}_1 \times \cdots \times \mathbf{A}_n)\,(\mathbf{B}_1 \times \cdots \times \mathbf{B}_n) = (\mathbf{A}_1\,\mathbf{B}_1) \times \cdots \times (\mathbf{A}_n\,\mathbf{B}_n)$. If $\mathbf{A}$ and $\mathbf{B}$ are invertible square matrices, $(\mathbf{A} \times \mathbf{B})^{-1} = \mathbf{A}^{-1} \times \mathbf{B}^{-1}$.

We begin by expliciting the SM in a specific case, and let $L = 3$. The input is simply identified as $\boldsymbol{a}_n = a_n$, with alphabet $\mathcal{A} = \mathcal{A}_M$. We then choose a valid state $\boldsymbol{s}_n$, and observe that $\sigma_m$ (or $z_m$) is not a sufficient information for determining (13). Hence, by inspection of (13), we let the state be

$$\boldsymbol{s}_n = \left(s_n^{(0)}, s_n^{(1)}, s_n^{(2)}\right) = (z_{n-3}, a_{n-2}, a_{n-1}) \tag{16}$$

as in the single–$h$ case. Then, from (11), the state update function is expressed as

$$\begin{cases} s_{n+1}^{(0)} = z_{n-2} = (z_{n-3} + r_{n-2}a_{n-2})_{2p} = \left(s_n^{(0)} + r_{n-2}s_n^{(1)}\right)_{2p} \\ s_{n+1}^{(1)} = a_{n-1} = s_n^{(2)} \\ s_{n+1}^{(2)} = a_n \ . \end{cases} \tag{17}$$

Therefore the state–transition has the form $\boldsymbol{s}_{n+1} = \boldsymbol{g}(\boldsymbol{s}_n, a_n, n)$ whose dependence on $n$ is due to (17), where $r_{n-2}$ depends on $n$. Moreover, the dependence is periodic in $n$ because of the periodicity of $r_n$. From (16) we see that the state set is $\mathcal{S} = \mathcal{N}_{2p} \times \mathcal{A}_M^2$. Finally, the output function is defined by (13), which is explicitly

$$\begin{aligned} \boldsymbol{b}_n &= W_{2p}^{z_{n-3}} \boldsymbol{w}_{a_{n-2}} \otimes \boldsymbol{w}_{a_{n-1}} \otimes \boldsymbol{w}_{a_n} \\ &= W_{2p}^{s_n^{(0)}} \boldsymbol{w}_{s_n^{(1)}} \otimes \boldsymbol{w}_{s_n^{(2)}} \otimes \boldsymbol{w}_{a_n} \ . \end{aligned} \tag{18}$$

This relation has the form $\boldsymbol{b}_n = \boldsymbol{h}\left(\boldsymbol{s}_n, a_n\right)$, where $\boldsymbol{h}(\cdot)$ is not time dependent.

In the general case we have

PROPOSITION 2 In a multi–$h$ CPM with alphabet size $M$, memory $L$ and rational modulation indexes $h_n = r_n/p$, $n = 0, 1, \ldots, N_h$, the states of the SM $\mathcal{M}_{\text{CPM}}$ are defined by

$$\boldsymbol{s}_n = \left(s_n^{(0)}, s_n^{(1)}, \ldots, s_n^{(L-1)}\right) = (z_{n-L}, a_{n-L+1}, \ldots, a_{n-1}) \tag{19}$$

and the state set is

$$\mathcal{S} = \mathcal{N}_{2p} \times \mathcal{A}_M^{L-1} \ . \tag{20}$$

The state–transition function $\boldsymbol{s}_{n+1} = \boldsymbol{g}(\boldsymbol{s}_n, a_n, n)$, periodic in $n$ with period $N_h$, is defined by

$$\begin{cases} s_{n+1}^{(0)} = \left(s_n^{(0)} + r_{n-L+1}s_n^{(1)}\right)_{2p} \\ s_{n+1}^{(1)} = s_n^{(2)} \\ \quad \vdots \\ s_{n+1}^{(L-2)} = s_n^{(L-1)} \\ s_{n+1}^{(L-1)} = a_n \ . \end{cases} \tag{21}$$

The output function $\boldsymbol{b}_n = \boldsymbol{h}\,(\boldsymbol{s}_n, a_n)$ is given by

$$\boldsymbol{b}_n = W_{2p}^{s_n^{(0)}} \boldsymbol{w}_{s_n^{(1)}} \otimes \cdots \otimes \boldsymbol{w}_{s_n^{(L-1)}} \otimes \boldsymbol{w}_{a_n} \,. \quad \square \tag{22}$$

The complexity of the SM $\mathcal{M}_{\text{CPM}}$ is essentially determined by the cardinality of the state set, given by

$$I = |\mathcal{S}| = 2p\, M^{L-1} \,. \tag{23}$$

The input alphabet is $\mathcal{A} = \mathcal{A}_M$. The length of the output words $\boldsymbol{b}_n$ is

$$|\mathcal{A}_M^L| = M^L \triangleq N \tag{24}$$

and their structure is

$$\begin{array}{l} W_{2p}^{s_n^{(0)}}\,[1 \quad 0 \quad 0 \quad \cdots \quad 0 \quad 0] \\ W_{2p}^{s_n^{(0)}}\,[0 \quad 1 \quad 0 \quad \cdots \quad 0 \quad 0] \\ \quad\vdots \\ W_{2p}^{s_n^{(0)}}\,[0 \quad 0 \quad 0 \quad \cdots \quad 0 \quad 1] \end{array}$$

so that the output alphabet $\mathcal{B}$ is a subset of $\mathcal{W}_{2p}^N$.

### C. *Statistical description for spectral analysis*

The target is the evaluation of the *average PSD* $\overline{R}_v(f)$ of the multi–$h$ CPM signal $v(t)$. In this analysis the following random processes are involved: the input data $a_n = a(nT)$, the state sequence $s_n = s(nT)$, the word sequence $\boldsymbol{b}_n = \boldsymbol{b}(nT)$, and the CPM signal $v(t)$, $t \in \mathbb{R}$.

The assumptions on which we base our analysis are four, namely

1) The input alphabet is an $M$-ary alphabet $\mathcal{A}_M$, with $M$ even, containing odd symbols.
2) The input data $\{a_n\}$ are *stationary* and *statistically independent*, with given probabilities $q_\alpha = \mathrm{P}\,[a_n = \alpha]$, $\alpha \in \mathcal{A}_M$.
3) The modulation indexes $h_n$ are rational, namely $h_n = r_n/p$, $n = 0, 1, \ldots, N_h - 1$, $p$ being the common denominator of the fractions and $r_n$ being integers.
4) The modulation index sequence $\{h_n\}$ is PTI of period $N_h$.

As a consequence of 4), the sequential machine $\mathcal{M}_{\text{CPM}}$ of Proposition 2 and the filter bank of (7a) are PTI. This makes all the processes $\boldsymbol{s}_n$, $\boldsymbol{b}_n$, and $v(t)$ (Fig. 2).

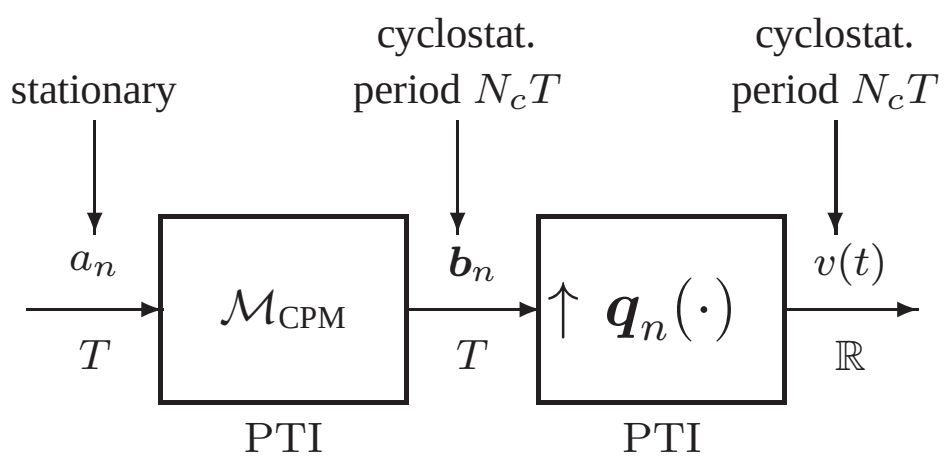

Fig. 2 – Representation of a multi–$h$ CPM modulator by a PTI SM followed by a PTI interpolator filterbank.

As we shall see, the period of cyclostationary $T_c$ depends not only on the period of the modulation indexes $N_h$, but also on their *parity*. In fact, the cyclostationarity period is $T_c = N_cT$ with

$$N_c = \begin{cases} N_h & \text{if } \sum_{n=0}^{N_h-1} r_n \text{ is even} \\ 2N_h & \text{if } \sum_{n=0}^{N_h-1} r_n \text{ is odd.} \end{cases} \tag{25}$$

Specifically, for a single-$h$ CPM with even $r_0$ we have $N_c = 1$, but for a odd $r_0$ it is $N_c = 2$.

The fundamental result for the statistical description of the above random processes is that, as a consequence of 4), the state sequence $\boldsymbol{s}_n$ is a *non–homogeneus* (i.e., non–stationary) Markov chain. Then, its full statistical specification can be obtained as a straightforward generalization of the results available for homogeneous Markov chains, providing the state absolute probabilities

$$p_n(\boldsymbol{i}) = \mathrm{P}\left[\boldsymbol{s}_n = \boldsymbol{i}\right] , \qquad \boldsymbol{i} \in \mathcal{S} \tag{26a}$$

and the state transition probabilities

$$\pi_n(\boldsymbol{i},\boldsymbol{j}) = \mathrm{P}\left[\boldsymbol{s}_{n+1} = \boldsymbol{i} | \boldsymbol{s}_n = \boldsymbol{j}\right] , \qquad \boldsymbol{i},\boldsymbol{j} \in \mathcal{S} \ . \tag{26b}$$

Both $p_n(\boldsymbol{i})$ and $\pi_n(\boldsymbol{i},\boldsymbol{j})$ are periodic of period $N_c$ in $n$. They are customarily collected in matrix form under the name of, respectively, the state absolute probability vector (APV) $\boldsymbol{p}_n$ and the transition probability matrix (TPM) $\boldsymbol{\pi}_n$ defined as

$$\boldsymbol{\pi}_n = \left[\pi_n(\boldsymbol{i},\boldsymbol{j})\right]_{\boldsymbol{i},\boldsymbol{j}\in\mathcal{S}} , \qquad \boldsymbol{p}_n = \left[p_n(\boldsymbol{i})\right]_{\boldsymbol{i}\in\mathcal{S}} \ . \tag{27}$$

### D. *Evaluation of the transition probability matrix*

For the evaluation of the TPM we can use the same technique of [15] with the modification concerning the time–dependence.

PROPOSITION 3 Let $\boldsymbol{e}_{\alpha,n} = [e_{\alpha,n}(\boldsymbol{i},\boldsymbol{j})]_{\boldsymbol{i},\boldsymbol{j}\in\mathcal{S}}$ be $I \times I$ binary matrices defined by the state–transition function $\boldsymbol{s}_{n+1} = \boldsymbol{g}_n(\boldsymbol{s}_n, \alpha)$ according to

$$e_{\alpha,n}(\boldsymbol{i},\boldsymbol{j}) = \begin{cases} 1 & \text{if } \boldsymbol{i} = \boldsymbol{g}_n(\boldsymbol{j},\alpha) \\ 0 & \text{if } \boldsymbol{i} \neq \boldsymbol{g}_n(\boldsymbol{i},\alpha) \end{cases} = \delta_{\boldsymbol{i},\,\boldsymbol{g}_n(\boldsymbol{j},\alpha)} \;. \tag{28}$$

We will call $\boldsymbol{e}_{\alpha,n}$ conditional transition matrices, since they express state transitions under the condition $a_n = \alpha$. Then

$$\boldsymbol{\pi}_n = \sum_{\alpha\in\mathcal{A}_M} q_\alpha\, \boldsymbol{e}_{\alpha,n} \,. \quad \square \tag{29}$$

Now, by use of (28) and the state–update function (21), the conditional transition matrices can be explicited as

$$\begin{aligned} \boldsymbol{e}_{\alpha,n} &= \left[\delta_{\boldsymbol{i},\,\boldsymbol{g}_n(\boldsymbol{j},\alpha)}\right]_{\boldsymbol{i},\boldsymbol{j}\in\mathcal{S}} \\ &= \left[\delta_{i_0,(j_0+r_{n-L+1}j_1)_{2p}}\, \delta_{i_1,j_2}\, \ldots\, \delta_{i_{L-2},j_{L-1}}\, \delta_{i_{L-1},\alpha}\right]_{i_0,j_0\in\mathcal{N}_{2p},\, i_1,j_1,\ldots,i_{L-1},j_{L-1}\in\mathcal{A}_M} \end{aligned} \tag{30}$$

where $\boldsymbol{i} = [i_0 \cdots i_{L-1}]^T$ and similarly for $\boldsymbol{j}$. The product of functions in (30) permits formulating the matrix $\boldsymbol{e}_{\alpha,n}$ in compact form through a Kronecker product by following a procedure similar to that leading to (13).

By preliminarily defining the single step ciclical shift matrix

$$\boldsymbol{D}_{2p} \triangleq \begin{bmatrix} & & & 1 \\ 1 & & & \\ & \ddots & & \\ & & 1 & \end{bmatrix} \qquad \text{matrix } 2p \times 2p \;, \tag{31}$$

which is a square matrix obtained by cyclically shifting the main diagonal to the left by $1$ position, we then have (see Appendix A for a proof)

PROPOSITION 4 The $I \times I$ conditional transition matrices $\boldsymbol{e}_{\alpha,n}$ for $L \geq 2$ are given by

$$\boldsymbol{e}_{\alpha,n} = \sum_{\beta\in\mathcal{A}_M} \boldsymbol{D}_{2p}^{r_{n-L+1}\beta} \otimes \boldsymbol{w}_\beta^T \otimes \boldsymbol{I}_{M^{L-2}} \otimes \boldsymbol{w}_\alpha \tag{32}$$

As a consequence, and by exploiting (29), the TPMs $\boldsymbol{\pi}_n$ give

$$\boldsymbol{\pi}_n = \sum_{\beta\in\mathcal{A}_M} \boldsymbol{D}_{2p}^{r_{n-L+1}\beta} \otimes \boldsymbol{w}_\beta^T \otimes \boldsymbol{I}_{M^{L-2}} \otimes \boldsymbol{q} \;. \tag{33}$$

The case $L = 1$ must be treated separately. We have

$$\boldsymbol{e}_{\alpha,n} = \boldsymbol{D}_{2p}^{r_{n-L+1}\alpha} \,, \qquad \boldsymbol{\pi}_n = \sum_{\beta\in\mathcal{A}_M} q_\beta\, \boldsymbol{D}_{2p}^{r_{n-L+1}\beta} \quad \square \tag{34}$$

### E. Evaluation of the absolute probability vector

With respect to the APV $\boldsymbol{p}_n$, the following update relation holds from the definition of $\boldsymbol{p}_n$ and $\boldsymbol{\pi}_n$,

$$\boldsymbol{p}_{n+1} = \boldsymbol{\pi}_n \, \boldsymbol{p}_n \; . \tag{35}$$

The interest is to identify that particular APV which is invariant to the update (35), that is

$$\boldsymbol{p}_n = \boldsymbol{\pi}_n \, \boldsymbol{p}_n \; ,$$

providing information on the limiting probabilities $\boldsymbol{p}_\infty$. Note that the APV we are looking for is an eigenvector of $\boldsymbol{\pi}_n$ with unit eigenvalue. As proved in Appendix B, such a vector of interest is

$$\boldsymbol{p}_n = \frac{1}{2p} \, \mathbf{1}_{2p} \otimes \underbrace{\boldsymbol{q} \otimes \cdots \otimes \boldsymbol{q}}_{L-1} \; . \tag{36}$$

## III. Classification of the Markov Chain

### A. Periodicity and reducibility properties

We have already seen that the TPMs $\boldsymbol{\pi}_n$ are periodic with period $N_h$. This makes the Markov chain being itself PTI. In addition, the Markov chain may further be *reducible*, that is if it is not possible to get to any state from any state or, equivalently, if some of the states cannot communicate between them. In the specific case of CPM, reducibility is assured by the rationale, explained in the following, and based upon the renewal law (11).

Recalling that $a_{m+1}$ is an odd number, we have the following prospect

$$\begin{array}{ll} r_{m+1} \text{ even} & \begin{cases} z_m \text{ even} & \to \quad z_{m+1} \text{ even} \\ z_m \text{ odd} & \to \quad z_{m+1} \text{ odd} \end{cases} \\ r_{m+1} \text{ odd} & \begin{cases} z_m \text{ even} & \to \quad z_{m+1} \text{ odd} \\ z_m \text{ odd} & \to \quad z_{m+1} \text{ even} \end{cases} \end{array}$$

Now, for convenience in the state set $\mathcal{S} = \mathcal{N}_{2p} \times \mathcal{A}_M^{L-1}$ we partition $\mathcal{N}_{2p}$ in its even and odd parts

$$\mathcal{N}_{2p}(+) = \{0, 2, \ldots, 2p-1\} \, , \qquad \mathcal{N}_{2p}(-) = \{1, 3, \ldots, 2p-1\} \; , \tag{37}$$

As a consequence, we arrange the matrix and vector indexes by displaying the even integers $\mathcal{N}_{2p}(+)$ first, followed by the odd integers $\mathcal{N}_{2p}(-)$, and find that the newly arranged matrices $\tilde{\boldsymbol{e}}_{\alpha,n}$ and $\tilde{\boldsymbol{\pi}}_n$ take a *block diagonal* form or a *block antidiagonal form*.

As an example, with $L = 1$, $p = 4$, $M = 2$ and $r_{n-L+1} = 3$, we find that $\tilde{\boldsymbol{\pi}}_n$ has the block antidiagonal form

| | 0 | 2 | 4 | 6 | 1 | 3 | 5 | 7 |
|---|---|---|---|---|---|---|---|---|
| 0 | | | | | 0 | $q_{-1}$ | $q_1$ | 0 |
| 2 | | | | | 0 | 0 | $q_{-1}$ | $q_1$ |
| 4 | | | | | $q_1$ | 0 | 0 | $q_{-1}$ |
| 6 | | | | | $q_{-1}$ | $q_1$ | 0 | 0 |
| 1 | 0 | 0 | $q_{-1}$ | $q_1$ | | | | |
| 3 | $q_1$ | 0 | 0 | $q_{-1}$ | | | | |
| 5 | $q_{-1}$ | $q_1$ | 0 | 0 | | | | |
| 7 | 0 | $q_{-1}$ | $q_1$ | 0 | | | | |

while for $r_{n-L+1} = 2$ we obtain the block diagonal form

| | 0 | 2 | 4 | 6 | 1 | 3 | 5 | 7 |
|---|---|---|---|---|---|---|---|---|
| 0 | 0 | $q_{-1}$ | 0 | $q_1$ | | | | |
| 2 | $q_1$ | 0 | $q_{-1}$ | 0 | | | | |
| 4 | 0 | $q_1$ | 0 | $q_{-1}$ | | | | |
| 6 | $q_{-1}$ | 0 | $q_1$ | 0 | | | | |
| 1 | | | | | 0 | $q_{-1}$ | 0 | $q_1$ |
| 3 | | | | | $q_1$ | 0 | $q_{-1}$ | 0 |
| 4 | | | | | 0 | $q_1$ | 0 | $q_{-1}$ |
| 7 | | | | | $q_{-1}$ | 0 | $q_1$ | 0 |

Note that submatrices are equal or closely related one each other by a simple circular shifts of the row (or column) order.

The result can be generalized by inspection of (21) and (26b). Specifically, for $\tilde{\boldsymbol{\pi}}_n$ we obtain the structure

$$\tilde{\boldsymbol{\pi}}_n = \begin{cases} \begin{bmatrix} \boldsymbol{F}_n & \boldsymbol{0} \\ \boldsymbol{0} & \boldsymbol{F}_n \end{bmatrix} & r_{n-L+1} \text{ even} \\ \begin{bmatrix} \boldsymbol{0} & \boldsymbol{D}_{I_0}^{I_0/p} \boldsymbol{G}_n \\ \boldsymbol{G}_n & \boldsymbol{0} \end{bmatrix} & r_{n-L+1} \text{ odd.} \end{cases} \tag{38}$$

where $\boldsymbol{F}_n$ and $\boldsymbol{G}_n$ are $I_0 \times I_0$ matrices with

$$I_0 = \frac{1}{2}\, I = p\, M^{L-1} \tag{39}$$

Note that all submatrices are themselves TPMs.

Moreover, submatices can be easily formulated as Kronecker products by recalling (33). For $r_{n-L+1}$ even we have

$$\boldsymbol{F}_n = \sum_{\beta\in\mathcal{A}_M} \boldsymbol{D}_{2p}^{r_{n-L+1}\beta}(+|+) \otimes \boldsymbol{w}_\beta^T \otimes \boldsymbol{I}_{M^{L-2}} \otimes \boldsymbol{q}$$

with $\boldsymbol{D}_{2p}^{r_{n-L+1}\beta}(+|+)$ the square matrix collecting the samples of $\boldsymbol{D}_{2p}^{r_{n-L+1}\beta}$ at even rows and even columns, while for $r_{n-L+1}$ odd it is

$$\boldsymbol{G}_n = \sum_{\beta\in\mathcal{A}_M} \boldsymbol{D}_{2p}^{r_{n-L+1}\beta}(-|+) \otimes \boldsymbol{w}_\beta^T \otimes \boldsymbol{I}_{M^{L-2}} \otimes \boldsymbol{q}$$

with $\boldsymbol{D}_{2p}^{r_{n-L+1}\beta}(-|+)$ the square matrix collecting the samples of $\boldsymbol{D}_{2p}^{r_{n-L+1}\beta}$ at odd rows and even columns. Note also that the correction term in (38) can be written as

$$\boldsymbol{D}_{I_0}^{I_0/p} = \boldsymbol{D}_p \otimes \boldsymbol{I}_{M^{L-1}} \;, \tag{40}$$

which restores the true modulo $2p$ operation when transiting from odd to even states. Incidentally, the following equivalences hold

$$\boldsymbol{D}_{I_0}^{I_0/p}\,\boldsymbol{F}_n = \boldsymbol{F}_n\,\boldsymbol{D}_{I_0}^{I_0/p}\,, \qquad \boldsymbol{D}_{I_0}^{I_0/p}\,\boldsymbol{G}_n = \boldsymbol{G}_n\,\boldsymbol{D}_{I_0}^{I_0/p}\;. \tag{41}$$

By now decomposing the state set $\mathcal{S} = \mathcal{N}_{2p} \times \mathcal{A}_M^{L-1}$ into the subsets

$$\mathcal{S}(\pm) = \mathcal{N}_{2p}(\pm) \times \mathcal{A}_M^{L-1} \;.$$

we can further illustrate the meaning of matrices in (38) by the following graph:

$$\begin{array}{ll} r_{n-L+1}\text{ even} & \begin{cases} \mathbf{s}_n \in \mathcal{S}(+) \xrightarrow{\;\boldsymbol{F}_n\;} \mathbf{s}_n \in \mathcal{S}(+) \\ \mathbf{s}_n \in \mathcal{S}(-) \xrightarrow{\;\boldsymbol{F}_n\;} \mathbf{s}_n \in \mathcal{S}(-) \end{cases} \\ r_{n-L+1}\text{ odd} & \begin{cases} \mathbf{s}_n \in \mathcal{S}(+) \xrightarrow{\;\boldsymbol{G}_n\;} \mathbf{s}_n \in \mathcal{S}(-) \\ \mathbf{s}_n \in \mathcal{S}(-) \xrightarrow{\;\boldsymbol{D}_{I_0}^{I_0/p}\boldsymbol{G}_n\;} \mathbf{s}_n \in \mathcal{S}(+) \end{cases} \end{array} \tag{42}$$

An equivalent illustration is given in Fig. 3.

We see that, with $r_{n-L+1}$ even, the state sets $\mathcal{S}(\pm)$ do not communicate: starting with $\boldsymbol{s}_n \in \mathcal{S}(+)$ the next state $\boldsymbol{s}_{n+1}$ is again in $\mathcal{S}(+)$, and this transition is governed by the TPM $\boldsymbol{F}_n$. Analogously, starting with $\mathrm{s}_n \in \mathcal{S}(-)$ the next state is $\boldsymbol{s}_{n+1} \in \mathcal{S}(-)$ with TPM $\boldsymbol{F}_n$. No transition is possible from $\mathcal{S}(+)$ into $\mathcal{S}(-)$ and from $\mathcal{S}(-)$ into $\mathcal{S}(+)$. For $r_{n-L+1}$ odd, the state sets $\mathcal{S}(\pm)$ do communicate deterministically: starting with $\boldsymbol{s}_n \in \mathcal{S}(+)$ the next state is $\boldsymbol{s}_{n+1} \in \mathcal{S}(-)$ with TPM $\boldsymbol{G}_n$, etc.

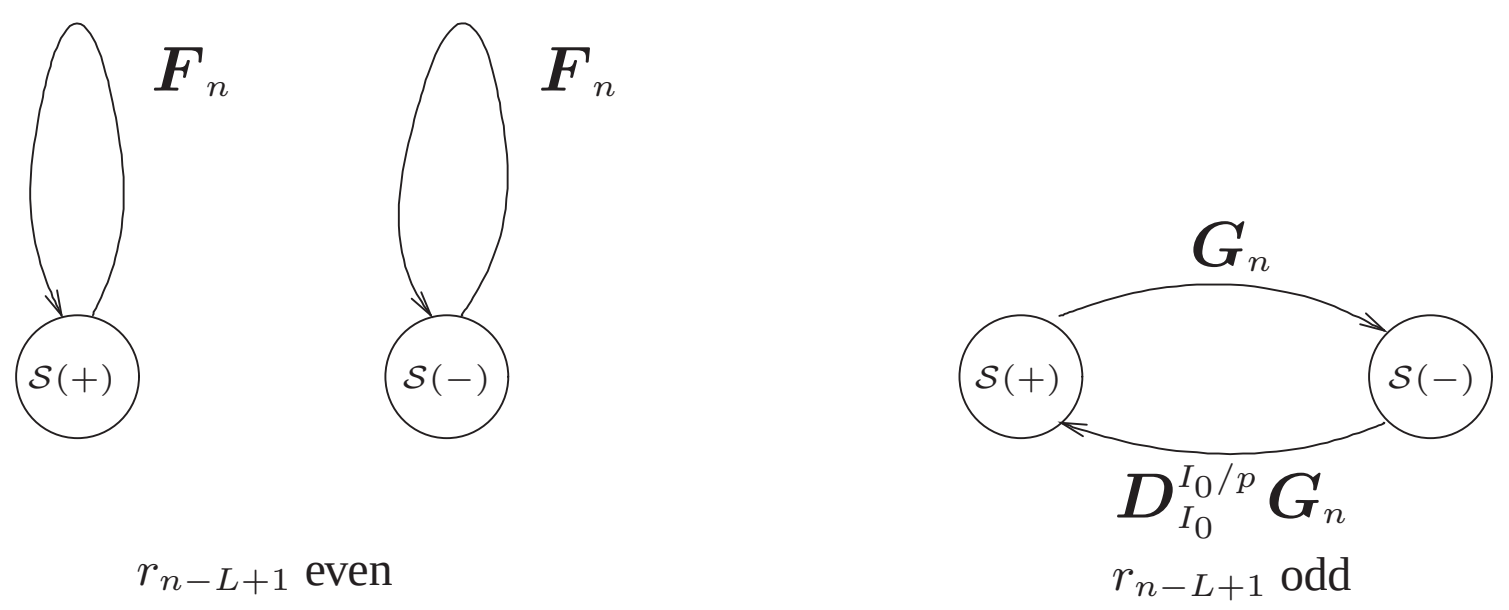

Fig. 3 – State transition graphs from classes $\mathcal{S}(+)$ and $\mathcal{S}(-)$.

## *B. Concluding remarks on Markov chain evaluation*

At this point we realize the fundamental role of the sequence $r_{n-L+1}$ of the *normalized* (to $2p$) modulation indexes. We start introducing two examples and conclude with the general case after.

EXAMPLE 1 Let $L = 1$, $N_h = 3$, $r_0$ even, $r_1$ and $r_2$ odd. We consider the state sequence $\boldsymbol{s}_n$ starting from $n = 0$. If $\boldsymbol{s}_0 \in \mathcal{S}(+)$, the transition is $\boldsymbol{s}_1 \in \mathcal{S}(+)$ with TPM $\boldsymbol{F}_0$, next the transitions are $\boldsymbol{s}_2 \in \mathcal{S}(-)$ with TPM $\boldsymbol{G}_1$ and $\boldsymbol{s}_3 \in \mathcal{S}(+)$ with TPM $\boldsymbol{D}_{I_0}^{I_0/p}\boldsymbol{G}_2$, etc. So we find the TPM sequence

$$\boldsymbol{F}_0\ ,\ \boldsymbol{G}_1\ ,\ \boldsymbol{D}_{I_0}^{I_0/p}\boldsymbol{G}_2\ ,\ \boldsymbol{F}_0\ ,\ \boldsymbol{G}_1\ ,\ \boldsymbol{D}_{I_0}^{I_0/p}\boldsymbol{G}_2\ ,\ \boldsymbol{F}_0\ ,\ \ldots$$

as illustrated at the top of Fig. 4.

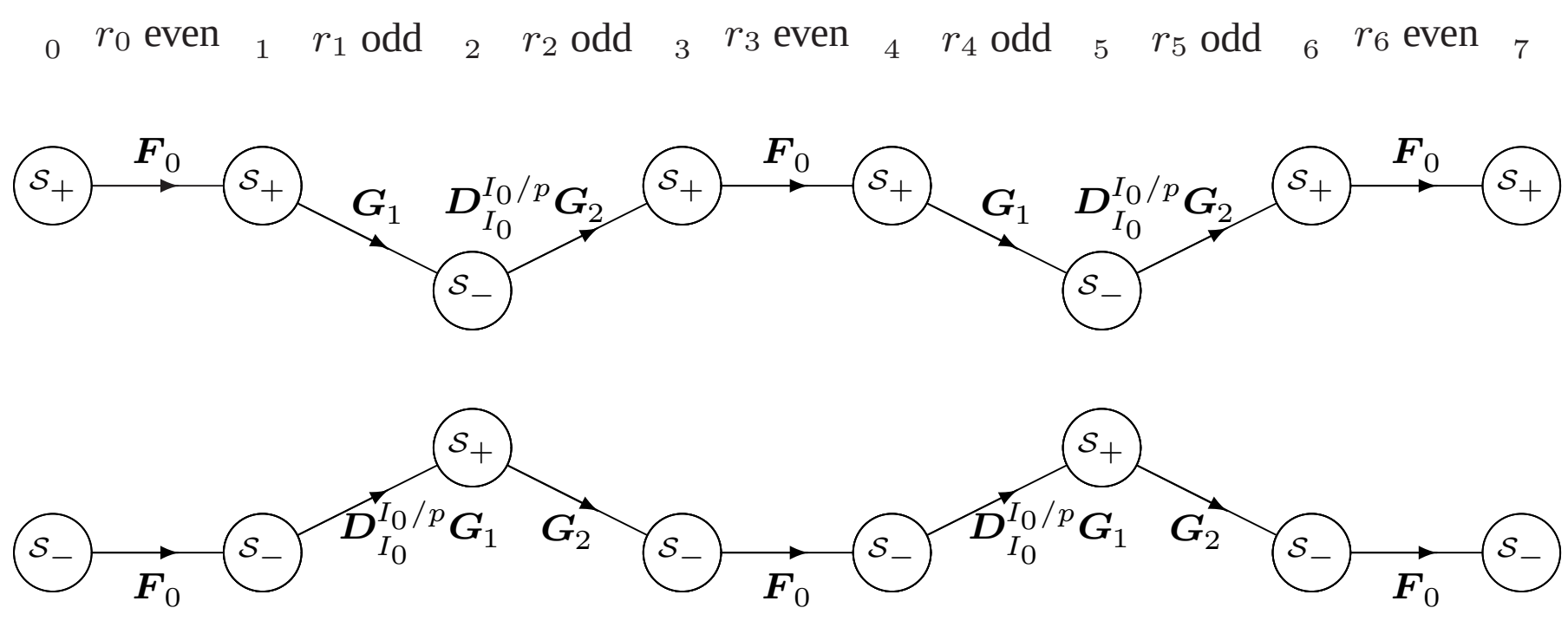

Fig. 4 – Trajectories of Example1.

Analogously, if $\boldsymbol{s}_0 \in \mathcal{S}(-)$ we find the complemetary sequence illustrated below in Fig. 4. In both cases the period is $N_c = N_h = 3$.

EXAMPLE 2 Now let $L = 1$ and $N_h = 3$, and suppose $r_0$ and $r_2$ even, $r_1$ odd. Depending on the state $\boldsymbol{s}_0$, we find the two trajectories illustrated in Fig. 5. In both case the period is $N_c = 2N_h = 6$.

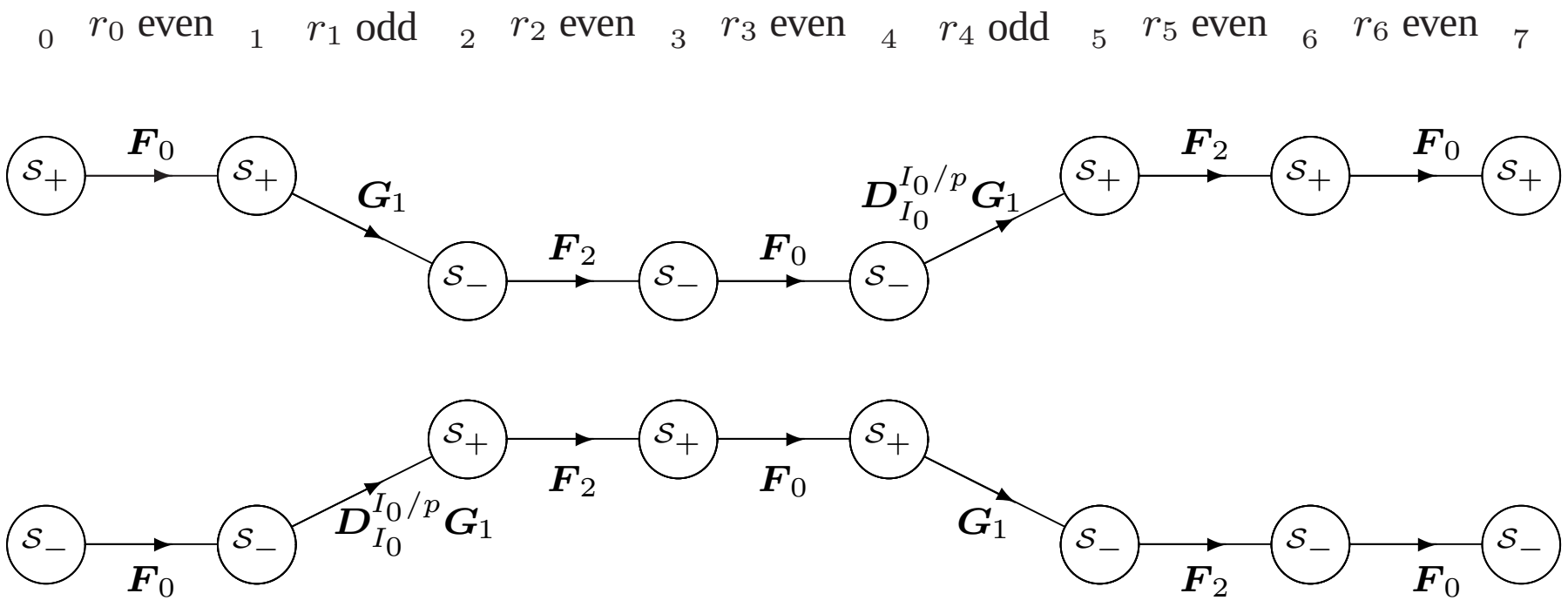

Fig. 5 – Trajectories of Example2.

EXAMPLE 3 Let $L = 1$, $N_c = 2$, $r_0$ be even, and $r_1$ be odd. If $\mathrm{s}_0 \in \mathcal{S}(+)$ the sequence of TPMs is illustrated in Fig.6 with period $N_c = 2N_h = 4$.

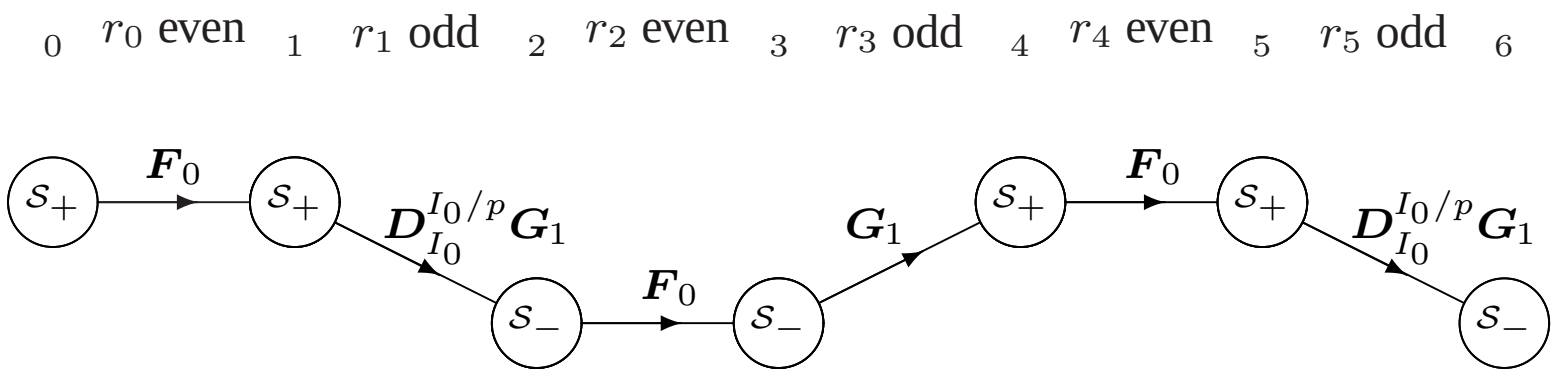

Fig. 6 – Trajectories of Example3.

From the above examples and from the general rules (42) (see also Fig.3), we see that, starting from $n = 0$, the Markov chain

$$\boldsymbol{s}_0\,, \qquad \boldsymbol{s}_1\,, \qquad \boldsymbol{s}_2\,, \ldots$$

has two distinct classes of trajectories $\mathcal{T}(\pm)$ which do not communicate each other anytime. The first class $\mathcal{T}(+)$ is determined by the condition $\boldsymbol{s}_0 \in \mathcal{S}(+)$ and the second class $\mathcal{T}(-)$ by the condition $\boldsymbol{s}_0 \in \mathcal{T}(-)$. We can extend these considerations to the bilateral chain

$$\ldots\,, \qquad \boldsymbol{s}_{-2}\,, \qquad \boldsymbol{s}_{-1}\,, \qquad \boldsymbol{s}_0\,, \qquad \boldsymbol{s}_1\,, \qquad \boldsymbol{s}_2\,, \ldots$$

which may be candidate for stationarity or cyclostationarity.

In the underlying probability space, we have to operate under one of the conditions[2]

$$\mathcal{C}_+ : \boldsymbol{s}_0 \in \mathcal{S}(+) \,, \qquad \mathcal{C}_- : \boldsymbol{s}_0 \in \mathcal{S}(-)$$

and correspondingly we find two distinct classes of trajectories:

$$\{\boldsymbol{s}_n\} \in \mathcal{T}(+) \quad \text{and} \quad \{\boldsymbol{s}_n\} \in \mathcal{T}(-) \;.$$

We claim that both classes can be modeled by a non–homogeneous irreducible Markov chain with TPMs $\boldsymbol{\pi}_n(+)$ and $\boldsymbol{\pi}(-)$, respectively. The cardinality of these conditional chains is $pM^{L-1}$, that is half the cardinality $2pM^{L-1}$ of the original unconditional Markov chain.[3]

The TPMs $\boldsymbol{\pi}_n(\pm)$ are obtained from the TPM $\boldsymbol{\pi}_n$ of the unconditioned Markov chain with the block partition and the rules seen above and now summarized.

PROPOSITION5 Let $\boldsymbol{\pi}_n$ be the TPM of the unconditioned Markov chain, which has the form (38) and period $N_c$. The period $N_c$ of $\boldsymbol{\pi}_n(\pm)$ depends on the sequence of the normalized modulation indexes $r_0, r_1, \ldots, r_{N_h-1}$, namely

$$N_c = \begin{cases} N_h & \text{if the number of the } r_n \text{ odd in a period } N_h \text{ is even} \\ 2N_h & \text{if the number of the } r_n \text{ odd in a period } N_h \text{ is odd.} \end{cases}$$

The $\boldsymbol{\pi}_n(+)$ in a period starts with $\boldsymbol{\pi}_0(+) = \boldsymbol{F}_0$ if $r_{-L+1}$ is even ($\boldsymbol{s}_1 \in \mathcal{S}(+)$), and with $\boldsymbol{\pi}_0(+) = \boldsymbol{G}_0$ if $r_{-L+1}$ is odd ($\boldsymbol{s}_1 \in \mathcal{S}(-)$). The rest is obtained recursively with the rule (42). Analogously $\boldsymbol{\pi}_n(-)$ starts with $\boldsymbol{\pi}_0(-) = \boldsymbol{F}_0$ if $r_{-L+1}$ is even ($\boldsymbol{s}_1 \in \mathcal{S}(-)$), and with $\boldsymbol{\pi}_0(-) = \boldsymbol{D}_{I_0}^{I_0/p}\,\boldsymbol{G}_0$ if $r_{-L+1}$ is odd ($\boldsymbol{s}_1 \in \mathcal{S}(+)$), and is obtained with the same rules. □

## IV. Alternative Representation of CPM Modulator

A promising approach to closed form PSD evaluation is given by reinterpreting the CPM modulator through the *polyphase decomposition* (PD) or serial-to-parallel conversion (S/P) of Fig. 7, where the input symbol sequence $\{a_n\}$ is turned into the word sequence

$$\boldsymbol{x}_n = \left[a_{nN_c}, a_{nN_c+1}, \cdots, a_{nN_c+N_c-1}\right] \tag{43}$$

which is itself a stationary sequence with alphabet $\mathcal{A}_M^{N_c}$.

[2]In practice, the condition is determined by the remote history of the modulator.

[3]The novelty (to be clarified) is that the state of each conditioned Markov chain is not unique but it changes from $\mathcal{S}(+)$ and $\mathcal{S}(-)$ in dependence of the modulation indexes. However, $\mathcal{T}(+)$ and $\mathcal{T}(-)$ are isomorphic.

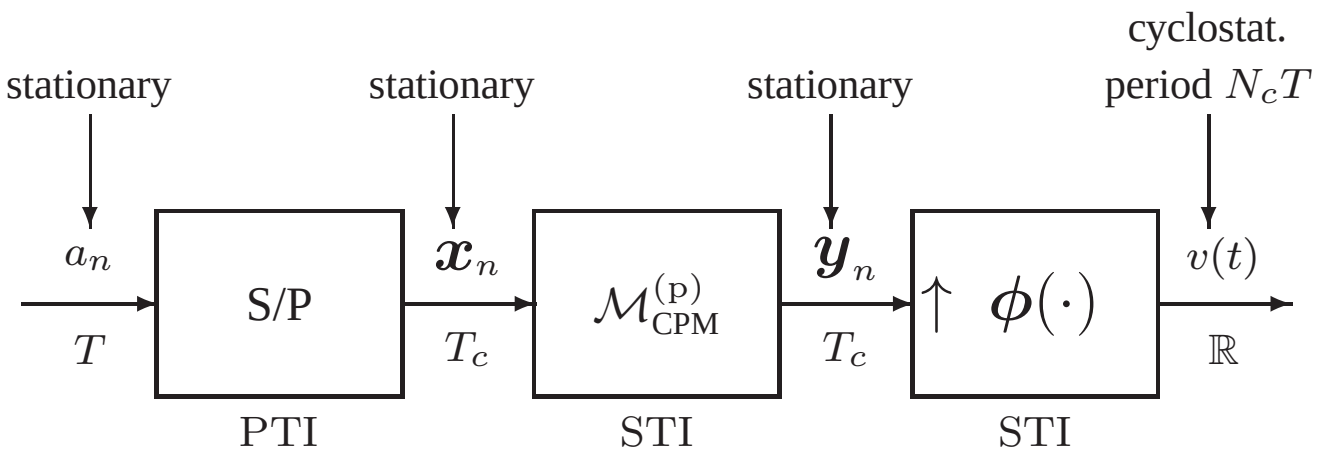

Fig. 7

Representation of a multi–$h$ CPM modulator by a S/P conversion followed by a time invariant SM and interpolator filter-bank.

The subsequent SM generating the output words $\{\boldsymbol{y}_n\}$ can then be formulated on an equivalent to (3), with the only difference that now the observation interval of interest becomes of length $T_c$ with the form $\mathcal{I}_n = [nT_c, nT_c + T_c)$. We then have

$$\alpha(t) = 2\pi \left[ \sum_{m=-\infty}^{nN_c-L} a_m h_m \frac{1}{2} + \sum_{m=nN_c-L+1}^{nN_c+N_c-1} a_m h_m \, \varphi(t - mT) \right], \qquad t \in \mathcal{I}_n \,, \tag{44}$$

so that the counterpart to (7) becomes

$$v(t) = \sum_n \boldsymbol{\phi}(t - nT)\, \boldsymbol{y}_n \tag{45}$$

where

$$\boldsymbol{y}_n = \sigma_{nN_c-L} \left[ \delta_{\boldsymbol{a}_n, \boldsymbol{\alpha}} \right]_{\boldsymbol{\alpha} \in \mathcal{A}_M^{L+N_c-1}}$$

$$\boldsymbol{\phi}(t) = \eta_{T_c}(t) \left[ \prod_{i=0}^{L-1} \exp\left( j2\pi h_{nN_c-i} \alpha_i \varphi(t + iT) \right) \prod_{i=1}^{N_c-1} \exp\left( j2\pi h_{nN_c+i} \beta_i \varphi(t - iT) \right) \right]_{\boldsymbol{\alpha} \in \mathcal{A}_M^{L+N_c-1}} \tag{45a}$$

Here, $\eta_{T_c}(t)$ is an indicator function active over the interval $[0, T_c)$, $\boldsymbol{a}_n = (a_{nN_c-L+1}, \ldots, a_{nN_c+N_c-1})$ collects the input data over a window of length $N_c+L-1$, $\boldsymbol{\alpha} = (\alpha_0, \alpha_1, \ldots, \alpha_{L-1}, \beta_1, \ldots \beta_{N_c-1}) \in \mathcal{A}_M^L$, and $N_0 = M^{L+N_c-1}$ is the cardinality of $\mathcal{A}_M^{L+N_c-1}$. So, $\boldsymbol{y}_n$ is a row vector of length $N_0$, while $\boldsymbol{\phi}(t)$ is a column vector of the same length.

Note that, unlike (7), $\boldsymbol{\phi}(t)$ is independent on $n$. Moreover, the second of (45a) clearly shows the newly required component driven by the symbols $\beta_i$. The formalization of a proper SM for (45) is now immediate by exploiting the results of Proposition 2.

PROPOSITION 6 In a multi–$h$ CPM with alphabet size $M$, memory $L$, rational modulation indexes $h_n = r_n/p$, $n = 0, 1, \ldots, N_h$, and time-invariance period $N_c$, the states of the time invariant SM $\mathcal{M}_{\text{CPM}}^{(\text{p})}$ (where p stands for parallel) are defined by

$$\boldsymbol{\sigma}_n = \left( \sigma_n^{(0)}, \sigma_n^{(1)}, \ldots, \sigma_n^{(L-1)} \right) = (z_{nN_c-L}, a_{nN_c-L+1}, \ldots, a_{nN_c-1}) \tag{46}$$

with a one-to-one relation to the PTI states (19) given by the sampling relation

$$\boldsymbol{\sigma}_n = \boldsymbol{s}_{nN_c} \; . \tag{47}$$

The output function $\boldsymbol{y}_n = \boldsymbol{h}\left(\boldsymbol{\sigma}_n, \boldsymbol{x}_n\right)$ is given by

$$\boldsymbol{y}_n = W_{2p}^{\sigma_n^{(0)}} \boldsymbol{w}_{\sigma_n^{(1)}} \otimes \cdots \otimes \boldsymbol{w}_{\sigma_n^{(L-1)}} \otimes \boldsymbol{w}_{x_n^{(0)}} \cdots \otimes \boldsymbol{w}_{x_n^{(N_c-1)}} \; . \quad \square \tag{48}$$

According to the ordering in (48) we can further attempt an equivalent Kronecker formulation of the interpolating filter bank $\boldsymbol{\phi}(t)$ as

$$\boldsymbol{\phi}(t) = \boldsymbol{\phi}_{-L+1}(t) \otimes \cdots \otimes \boldsymbol{\phi}_{N_c-1}(t) \tag{49}$$

where

$$\boldsymbol{\phi}_i(t) = \eta_{T_c}(t) \sum_{\alpha \in \mathcal{A}_m} \boldsymbol{w}_\alpha \exp\left(j2\pi h_i \alpha \varphi(t - iT)\right) \tag{49a}$$

In addition, by exploiting the sampling relationship (47), the statistical description of the newly introduced SM immediately follows from Proposition 3 and Proposition 4, and from general properties of Markov chains. We have

PROPOSITION 7 The $I \times I$ conditional transition matrices $\boldsymbol{E}_{\boldsymbol{\alpha},n}$ for $\boldsymbol{\alpha} = [\alpha_0, \ldots, \alpha_{N_c-1}]^T$ and $L \geq 2$ are defined through the matrix product

$$\boldsymbol{E}_{\boldsymbol{\alpha}} = \boldsymbol{e}_{\alpha_{N_c-1}, nN_c} \cdots \boldsymbol{e}_{\alpha_1, nN_c+1} \, \boldsymbol{e}_{\alpha_0, nN_c} \tag{50}$$

with $\boldsymbol{e}_{\alpha,n}$ as defined in (32). In turn, the TPM $\boldsymbol{\Pi}$ is built as the TPM matrix product

$$\boldsymbol{\Pi} = \boldsymbol{\pi}_{nN_c+N_c-1} \cdots \boldsymbol{\pi}_{nN_c+1} \, \boldsymbol{\pi}_{nN_c} \tag{51}$$

where $\boldsymbol{\pi}_n$ is given by (33). They both are independent on $n$. Observe that, for $L = 1$ expressions (34) must be used in place of (32) and (33). $\square$

As a consequence of (38) and Proposition 5, we also have

PROPOSITION 8 By using the state ordering (37), the TPM matrix $\boldsymbol{\Pi}$ takes the block diagonal form

$$\tilde{\boldsymbol{\Pi}} = \begin{bmatrix} \boldsymbol{\Pi}(+) & \boldsymbol{0} \\ \boldsymbol{0} & \boldsymbol{\Pi}(-) \end{bmatrix} \tag{52}$$

where $\mathbf{\Pi}(+)$ contains the samples of $\mathbf{\Pi}$ at even rows and even columns, while $\mathbf{\Pi}(-)$ contains the samples of $\mathbf{\Pi}$ at odd rows and odd columns. Moreover

$$\mathbf{\Pi}(\pm) = \boldsymbol{\pi}_{N_c-1}(\pm) \cdots \boldsymbol{\pi}_1(\pm)\, \boldsymbol{\pi}_0(\pm)\,, \qquad \mathbf{\Pi}(+) = \mathbf{\Pi}(-) \tag{53}$$

the second equivalence being assured by the permutation property (41). □

Incidentally, an equivalent to Proposition 8 holds for state-transition matrices $\boldsymbol{E}_{\boldsymbol{\alpha}}$, for which we have

$$\tilde{\boldsymbol{E}}_{\boldsymbol{\alpha}} = \begin{bmatrix} \boldsymbol{E}_{\boldsymbol{\alpha}}(+) & \mathbf{0} \\ \mathbf{0} & \boldsymbol{E}_{\boldsymbol{\alpha}}(-) \end{bmatrix}, \qquad \boldsymbol{E}_{\boldsymbol{\alpha}}(+) = \boldsymbol{E}_{\boldsymbol{\alpha}}(-)\ . \tag{54}$$

As a remark, we observe that the product formulation of (50), (51), and (53) could be restated through more direct expressions. However, this gives minor insights that are of no direct interest to spectral evaluation.

Finally, the APVs of interest, separately for the even and odd trajectories $\mathcal{T}(+)$ and $\mathcal{T}(-)$, can be formulated starting from (36) to give

$$\overset{\infty}{\boldsymbol{p}} = \mathbf{\Pi}(\pm)\, \overset{\infty}{\boldsymbol{p}}\,, \qquad \overset{\infty}{\boldsymbol{p}} = \frac{1}{p}\, \mathbf{1}_p \otimes \underbrace{\boldsymbol{q} \otimes \cdots \otimes \boldsymbol{q}}_{L-1} \tag{55}$$

Incidentally, since both trajectories $\mathcal{T}(\pm)$ identify irreducible Markov chains, the following limit property is known to hold (e.g. see [17])

$$\overset{\infty}{\mathbf{\Pi}}(\pm) \triangleq \lim_{k\to\infty} \mathbf{\Pi}(\pm)^k = \overset{\infty}{\boldsymbol{p}}\, \mathbf{1}_{I/2}^T \tag{56}$$

that is the limit probability of being in a state is independent on the initial state.

## V. Closed Form Spectral Evaluation

### A. Generalities

We follow the PD approach depicted in Fig. 7, which is more simple and permits the direct use of the theory of [15]. In fact, the reformulation of the SM in a time invariant form removes the PTI on the output sequence $\{\boldsymbol{y}_n\}$ which, unlike $\{\boldsymbol{b}_n\}$, is stationary. Once evaluated the correlation

$$\boldsymbol{r}_{\boldsymbol{y}}(kT_c) = \mathrm{E}\left[\boldsymbol{y}_m\, \boldsymbol{y}_{m-k}^*\right] \tag{57}$$

and the corresponding PSD

$$\boldsymbol{R}_{\boldsymbol{y}}(f) = T_c \sum_k \boldsymbol{r}_{\boldsymbol{y}}(kT_c)\, e^{-j2\pi f k T_c} \tag{58}$$

the output (average) PSD is simply given by

$$\overline{R}_v(f) = \boldsymbol{\Phi}(f)\, \boldsymbol{R}_{\boldsymbol{y}}(f)\, \boldsymbol{\Phi}^*(f) \tag{59}$$

with $\boldsymbol{\Phi}(f) = \int_{-\infty}^{+\infty} \boldsymbol{\phi}(t) e^{-j2\pi f t} dt$ the Fourier transform of $\boldsymbol{\phi}(t)$. The result (59) is a straightforward generalization of a widely known property of the interpolating filter to an interpolating filter bank, whose proof can be found in [ ???].

A fundamental role is played by the possible *presence of lines* (delta functions) in the SDs. In general, $\boldsymbol{R}_{\boldsymbol{y}}(f)$ is given by the sum of a series, which is not convergent. The technique to handle this divergency is the separation of the correlation $\boldsymbol{r}_{\boldsymbol{y}}(kT_c)$ into a *continous* part, $\boldsymbol{r}_{\boldsymbol{y}}^{(c)}(kT_c)$, and a *discrete* part, $\boldsymbol{r}_{\boldsymbol{y}}^{(d)}(kT_c)$. The latter is related to the asymptotic behavior of the correlation, and ultimately to the state transition probabilities (55). The PSDs will be henceforth explicitely written as

$$\boldsymbol{R}_{\boldsymbol{y}}(f) = \boldsymbol{R}_{\boldsymbol{y}}^{(c)}(f) + \boldsymbol{R}_{\boldsymbol{y}}^{(d)}(f) \tag{60}$$

and

$$\overline{R}_v(f) = \overline{R}_v^{(c)}(f) + \overline{R}_v^{(d)}(f) \ . \tag{61}$$

We shall see that, when present, the spectral lines occurs at the frequencies multiple of $F_c = 1/T_c$.

### B. New notation

We preliminarily need to assess some notation according to [15]. Let the output symbols $\boldsymbol{y}_n$ be organized in matrices such that $\boldsymbol{Y}_{\boldsymbol{x}}$ collects (in its columns) the output words corresponding to the input word $\boldsymbol{x}$. Evidently, being $I$ the number of states, such words are in number of $I$, and matrix $\boldsymbol{Y}_{\boldsymbol{x}}$ is thus a $N \times I$ matrix. By exploiting the first of (45a) and (48), in Kronecker form it is (see Appendix C)

$$\boldsymbol{Y}_{\boldsymbol{x}} = \boldsymbol{V}_{2p} \otimes \boldsymbol{I}_{M^{L-1}} \otimes \boldsymbol{w}_{x_0} \otimes \cdots \otimes \boldsymbol{w}_{x_{N_c-1}} \tag{62}$$

where

$$\boldsymbol{V}_{2p} = [W_{2p}^j]_{j \in \mathcal{N}_{2p}} = [1, W_{2p}, W_{2p}^2, \cdots, W_{2p}^{2p-1}] \ . \tag{62a}$$

is a row vector. To discriminate between trajectories $\mathcal{T}(\pm)$, we can further introduce the notation

$$\boldsymbol{Y_x}(+) = \boldsymbol{V}_p \otimes \boldsymbol{I}_{M^{L-1}} \otimes \boldsymbol{w}_{x_0} \otimes \cdots \otimes \boldsymbol{w}_{x_{N_c-1}}, \qquad \boldsymbol{Y_x}(-) = W_{2p}\, \boldsymbol{Y_x}(+) . \tag{63}$$

### C. Presence of spectral lines in $\boldsymbol{R_y}$

Spectral lines depend on the limit behavior of the Markow chain, specifically they are related to the limit mean value of $\boldsymbol{y}_n$, namely

$$\boldsymbol{m_y} = \lim_{n\to\infty} \mathrm{E}\left[\boldsymbol{y}_n\right] , \tag{64}$$

where APVs (55) hold. Now, the mean (64) can be evaluated as

$$\boldsymbol{m_y}(\pm) = \sum_{\boldsymbol{x}} \mathrm{P}\left[\boldsymbol{x}_\infty = \boldsymbol{x}\right] \boldsymbol{Y_x}(\pm)\, \overset{\infty}{\boldsymbol{p}} \tag{65}$$

from which we have the following general result (the proof is reported in Appendix D)

PROPOSITION 9 When $p > 1$, no spectral lines occurr in CPM and we have

$$\boldsymbol{R}_{\boldsymbol{y}}^{(d)}(f) = \boldsymbol{0}\,, \qquad p > 1\,. \tag{66a}$$

Conversely, when $p = 1$, that is for integer modulation factors, spectral lines are found and it is

$$\boldsymbol{R_y}^{(d)}(f) = \boldsymbol{r_y}^{(d)} \sum_k \delta(f - kF_c)\,, \qquad p = 1 \tag{66b}$$

where

$$\boldsymbol{r_y}^{(d)} = \underbrace{[\boldsymbol{q}\,\boldsymbol{q}^*] \otimes \cdots \otimes [\boldsymbol{q}\,\boldsymbol{q}^*]}_{N_c+L-1} \tag{67}$$

and $F_c = 1/T_c$. □

We note that Proposition 9 is in accordance to the results of the literature on single-$h$ CPM [18].

### D. Continuous spectrum evaluation of $\boldsymbol{R_y}$

In order to evaluate the continuous part of the spectrum we preliminarily need to further explicit the correlation $\boldsymbol{r_y}$ in (57) following the development in [15].

For the value at $nT_c = 0$ (autocorrelation) we have (see (36) in [15])

$$\boldsymbol{r_y}(0) = \mathrm{E}\left[\boldsymbol{y}_m\, \boldsymbol{y}_m^*\right] = \sum_{\boldsymbol{x}} \mathrm{P}\left[\boldsymbol{x}_m = \boldsymbol{x}\right]\, \boldsymbol{Y_x}(\pm)\, \mathrm{diag}(\overset{\infty}{\boldsymbol{p}})\, \boldsymbol{Y_x}^*(\pm)\,. \tag{68}$$

where, by exploiting the Kronecker product expressions (55) and (63), and after some little algebra, we have

$$\boldsymbol{r}_{\boldsymbol{y}}(0) = \underbrace{\mathrm{diag}(\boldsymbol{q}) \otimes \cdots \otimes \mathrm{diag}(\boldsymbol{q})}_{N_c+L-1} \,. \tag{69}$$

For the values at $n \neq 0$, we better distinguish between positive and negative values. For $n > 0$ we have (see (37)-(39) in [15])

$$\boldsymbol{r}_{\boldsymbol{y}}(nT_s) = \boldsymbol{C}_2(\pm)\,\boldsymbol{\Pi}(\pm)^{n-1}\,\boldsymbol{C}_1(\pm)\,, \qquad n > 0 \tag{70}$$

where

$$\begin{aligned} \boldsymbol{C}_1(\pm) &= \sum_{\boldsymbol{x}} \mathrm{P}\left[\boldsymbol{x}_m = \boldsymbol{x}\right] \boldsymbol{E}_{\boldsymbol{x}}(\pm)\,\mathrm{diag}(\overset{\infty}{\boldsymbol{p}})\,\boldsymbol{Y}_{\boldsymbol{x}}{}^*(\pm) \\ \boldsymbol{C}_2(\pm) &= \sum_{\boldsymbol{x}} \mathrm{P}\left[\boldsymbol{x}_m = \boldsymbol{x}\right] \boldsymbol{Y}_{\boldsymbol{x}}(\pm)\,. \end{aligned} \tag{70a}$$

Instead, for $n < 0$ we can exploit the relation

$$\boldsymbol{r}_{\boldsymbol{y}}(-nT_s) = \boldsymbol{r}_{\boldsymbol{y}}{}^*(nT_s)\,. \tag{71}$$

Observe that, the correlation samples of the two trajectories coincide since $\boldsymbol{E}_{\boldsymbol{x}}(+) = \boldsymbol{E}_{\boldsymbol{x}}(-)$ and $\boldsymbol{\Pi}(+) = \boldsymbol{\Pi}(-)$ (see (53) and (54)), while the $W_{2p}$ difference between $\boldsymbol{Y}_{\boldsymbol{x}}(+)$ and $\boldsymbol{Y}_{\boldsymbol{x}}(-)$ (see (63)) is removed by the complex conjugation in $\boldsymbol{C}_1$.

In addition, we must take into account for the limit value $\boldsymbol{r}_{\boldsymbol{y}}{}^{(d)}$, which (when not null) is responsible of spectral lines. According to (40) in [15] we can write

$$\boldsymbol{r}_{\boldsymbol{y}}{}^{(d)} = \boldsymbol{m}_{\boldsymbol{y}}(\pm)\,\boldsymbol{m}_{\boldsymbol{y}}{}^*(\pm) = \boldsymbol{C}_2(\pm)\,\overset{\infty}{\boldsymbol{\Pi}}\,\boldsymbol{C}_1(\pm)\,. \tag{72}$$

As the reader can verify by exploiting (56), this expression is not in contrast with the formulation in (67).

Now, the continuous spectrum becomes

$$\boldsymbol{R}_{\boldsymbol{y}}{}^{(c)}(f) = T_c \sum_n \left(\boldsymbol{r}_{\boldsymbol{y}}(nT_c) - \boldsymbol{r}_{\boldsymbol{y}}{}^{(d)}\right) e^{-j2\pi f n T_c} \tag{73}$$

and, by then relying on the symmetry in (71) and on the results of [15], the PSD can be written as

$$\begin{aligned} \boldsymbol{R}_{\boldsymbol{y}}{}^{(c)}(f) &= T_c \left(\boldsymbol{r}_{\boldsymbol{y}}(0) - \boldsymbol{r}_{\boldsymbol{y}}{}^{(d)}\right) + \\ &\quad + 2T_c\,\Re\left[\boldsymbol{C}_2(\pm)\left(\boldsymbol{I}_{I_0} - \overset{\infty}{\boldsymbol{\Pi}}\right)\left(e^{j2\pi f T_c}\boldsymbol{I}_{I_0} - \boldsymbol{\Pi}(\pm) + \overset{\infty}{\boldsymbol{\Pi}}\right)^{-1} \boldsymbol{C}_1(\pm)\right]\,. \end{aligned} \tag{74}$$

Note that this is a closed-form, unlike the results available in CPM literature (e.g., see [9], [13]) where the PSD is evaluated through direct numerical evaluation of the series (73).

We also underline that the matrix inversion in (74) can be circumvented. The details can be found in [15], while here we report the final result. In particular, the inversion of an $I \times I$ matrix $(\lambda\mathbf{I} - \mathbf{F})$, with $\lambda = e^{j2\pi f T_c}$, can be performed by

$$(\lambda\mathbf{I} - \mathbf{F})^{-1} = \frac{\sum_{k=0}^{I-1} \boldsymbol{G}_k\, \lambda^{I-1-k}}{d(\lambda)}\,, \qquad \boldsymbol{G}_k = \sum_{m=0}^{k} d_{k-m}\, \boldsymbol{F}^m$$

where

$$d(x) = \mathrm{trace}\Big(x\mathbf{I} - \mathbf{F}\Big) = \sum_{k=0}^{I} d_k\, x^{I-k}\ .$$

*E. Power spectral density of* $v(t)$

The derivation of the power spectral density of $v(t)$ can now be straightforwardly obtained by use of (59). For ease of computational evaluation, here we introduce the matices

$$\boldsymbol{V}_{\boldsymbol{x}}(f) = \boldsymbol{\Phi}(f)\, \boldsymbol{Y}_{\boldsymbol{x}}(+) \tag{75}$$

which are row matrices of dimension $1 \times I_0$, namely $\boldsymbol{V}_{\boldsymbol{x}}(f) = [V_{\boldsymbol{x},\boldsymbol{s}}(f)]_{\boldsymbol{s}\in\mathcal{S}(+)}$, whose element $V_{\boldsymbol{x},\boldsymbol{s}}(f)$ can be obtained from (see also (45a))

$$\begin{aligned} v_{\boldsymbol{x},\boldsymbol{s}}(t) = \eta_{T_c}(t)\, W_{2p}^{s^{(0)}} \prod_{i=1}^{L-1} \exp\Big(j2\pi\, h_{nN_c-L+i}\, s^{(i)}\, \varphi(t+(L-i)T)\Big)\ \cdot \\ \cdot \prod_{i=0}^{N_c-1} \exp\Big(j2\pi\, h_{nN_c+i}\, y^{(i)} \varphi(t-iT)\Big) \end{aligned} \tag{75a}$$

through an ordinary Fourier transform, that is $v_{\boldsymbol{x},\boldsymbol{s}}(t) \xrightarrow{\mathcal{F}} V_{\boldsymbol{x},\boldsymbol{s}}(f)$.

So, in the general case of non-integer modulation factors ($p > 1$) from (68) and (74) we obtain

$$\begin{aligned} &\overline{R}_v^{(d)}(f) = 0 \\ &\overline{R}_v^{(c)}(f) = T_c\, K_0(f) + 2T_c\, \Re\Big[\boldsymbol{K}_2(f)\, \Big(\boldsymbol{I}_{I_0} - \overset{\infty}{\boldsymbol{\Pi}}\Big) \Big(e^{j2\pi f T_c} \boldsymbol{I}_{I_0} - \boldsymbol{\Pi}(\pm) + \overset{\infty}{\boldsymbol{\Pi}}\Big)^{-1} \boldsymbol{K}_1(f)\Big] \end{aligned} \tag{76}$$

where

$$\begin{aligned} K_0(f) &= \sum_{\boldsymbol{x}} \mathrm{P}\left[\boldsymbol{x}_m = \boldsymbol{x}\right] \boldsymbol{V}_{\boldsymbol{x}}(f)\, \mathrm{diag}(\overset{\infty}{\boldsymbol{p}})\, \boldsymbol{V}_{\boldsymbol{x}}{}^*(f) \\ \boldsymbol{K}_1(f) &= \boldsymbol{C}_1(+)\, \boldsymbol{\Phi}^*(f) = \sum_{\boldsymbol{x}} \mathrm{P}\left[\boldsymbol{x}_m = \boldsymbol{x}\right] \boldsymbol{E}_{\boldsymbol{x}}(+)\, \mathrm{diag}(\overset{\infty}{\boldsymbol{p}})\, \boldsymbol{V}_{\boldsymbol{x}}{}^*(f) \\ \boldsymbol{K}_2(f) &= \boldsymbol{\Phi}(f)\, \boldsymbol{C}_2(+) = \sum_{\boldsymbol{x}} \mathrm{P}\left[\boldsymbol{x}_m = \boldsymbol{x}\right] \boldsymbol{V}_{\boldsymbol{x}}(f) \end{aligned} \tag{76a}$$

In the very special case of integer modulation factors ($p = 1$), we recall (67) to write the spectral lines as

$$\overline{R}_v^{(d)}(f) = \sum_k |\boldsymbol{K}_2(kF_c)\,\overset{\infty}{\boldsymbol{p}}|^2 \delta(f - kF_c) \tag{77}$$

while the continuous part of the spectrum in (76) needs a correction factor

$$\Delta\overline{R}_v^{(d)}(f) = -|\boldsymbol{K}_2(f)\,\overset{\infty}{\boldsymbol{p}}|^2 \tag{78}$$

according to (74).

## VI. Examples

In this section we give some examples of PSDs evaluated through (76). We begin by showing some results for *full response* CPM signals (i.e. with $L = 1$) in Fig. 8. All plots refer to a *continuous phase frequency shift keying* (CPFSK) employing a phase response

$$\varphi(t) = \begin{cases} 0 & , t < 0 \\ \frac{t}{2LT} & , 0 \le t < LT \\ \frac{1}{2} & , t \ge LT \end{cases} \tag{79}$$

and for different choices of the multi-$h$ sequence factors. Examples were taken from [9] and [10].

A further example of full response signaling taken from [9] is shown in Fig. 9, where CPFSK is compared to *raised cosine* (RC) signaling

$$\varphi(t) = \begin{cases} 0 & , t < 0 \\ \frac{t}{2LT} - \frac{1}{4\pi}\sin\left(2\pi\,\frac{t}{LT}\right) & , 0 \le t < LT \\ \frac{1}{2} & , t \ge LT \end{cases} \tag{80}$$

A final example is given for the partial response multi-$h$ CPM signaling formats of [12] in Fig. 10. The figure shows two RC formats, and one binary *Gaussian minimum shift keying* (GMSK) format with $L = 4$, having a phase response

$$\varphi(t) = \hat{\Phi}\Big(K\left(\tfrac{t}{T} - \tfrac{3}{2}\right)\Big) - \hat{\Phi}\Big(K\left(\tfrac{t}{T} - \tfrac{5}{2}\right)\Big), \qquad \hat{\Phi}(a) = a\,\Phi(a) + \frac{1}{\sqrt{2\pi}}e^{-\frac{1}{2}t^2} \tag{81}$$

where $K = \pi/(2\sqrt{\ln 2})$, and $\Phi(a) = \frac{1}{\sqrt{2\pi}}\int_{-\infty}^{a} e^{-\frac{1}{2}t^2}dt$ is the Gaussian normalized cumulative distribution function.

## VII. Conclusions

## Acknowledgements

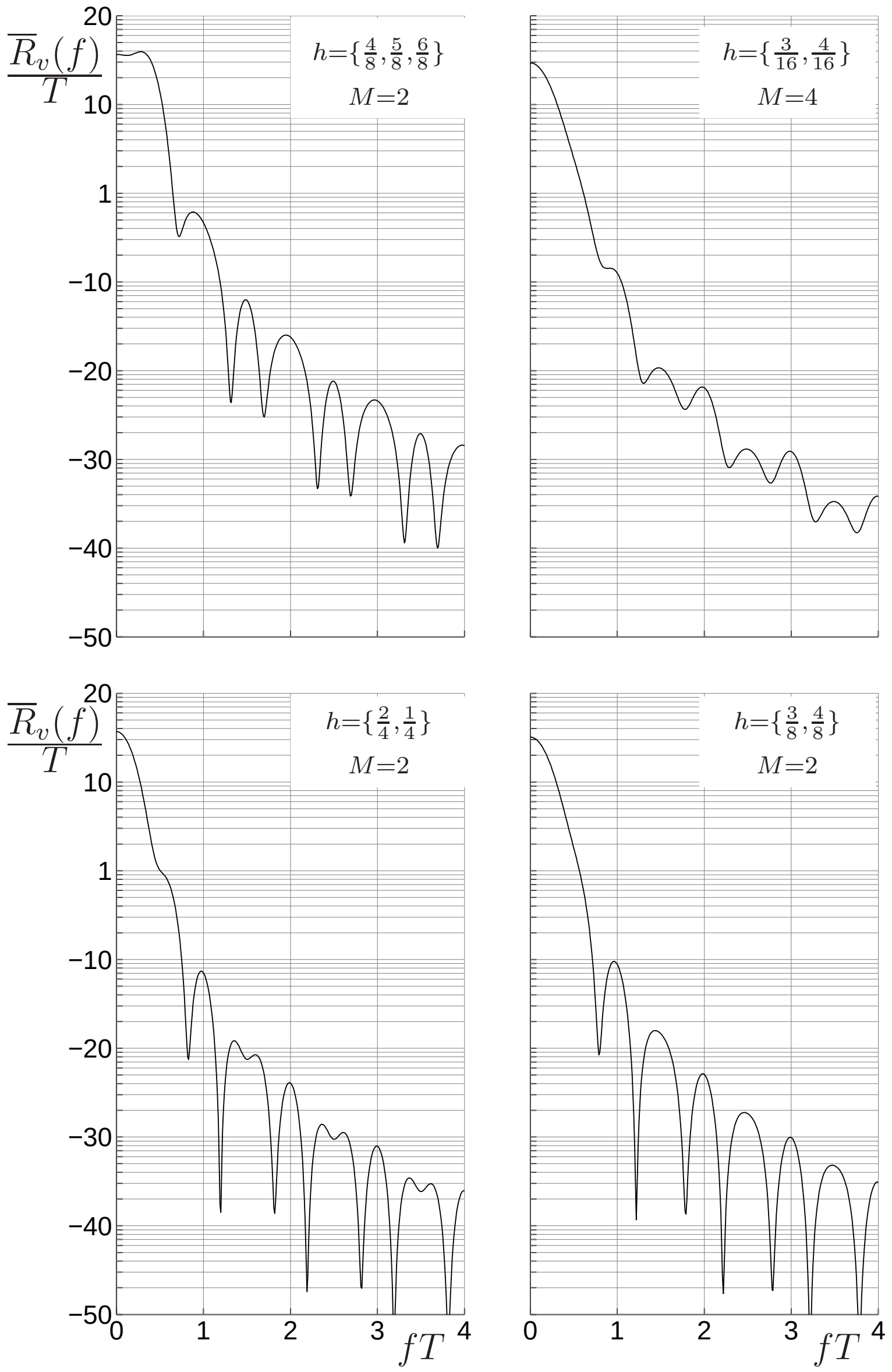

Fig. 8 – Normalised PSD plots of full response CPFSK, for different multi-$h$ sequences, and for equally likely input symbols.

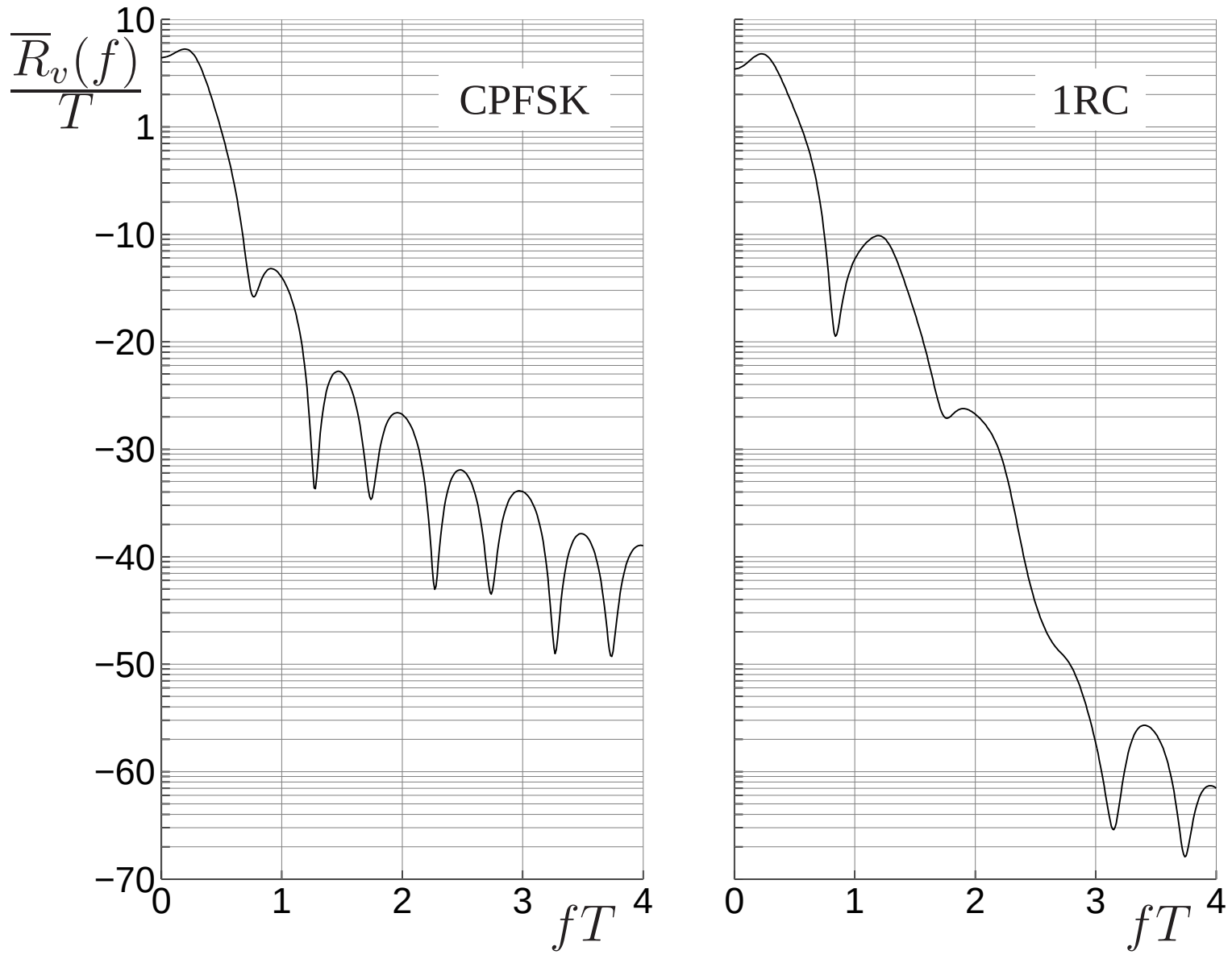

Fig. 9 – Normalised PSD plots for multi-$h$ sequence $\{\frac{4}{9}, \frac{6}{9}\}$, equally likely input symbols, $M = 2$ and $L = 1$. Plots show full response CPFSK and 1RC.

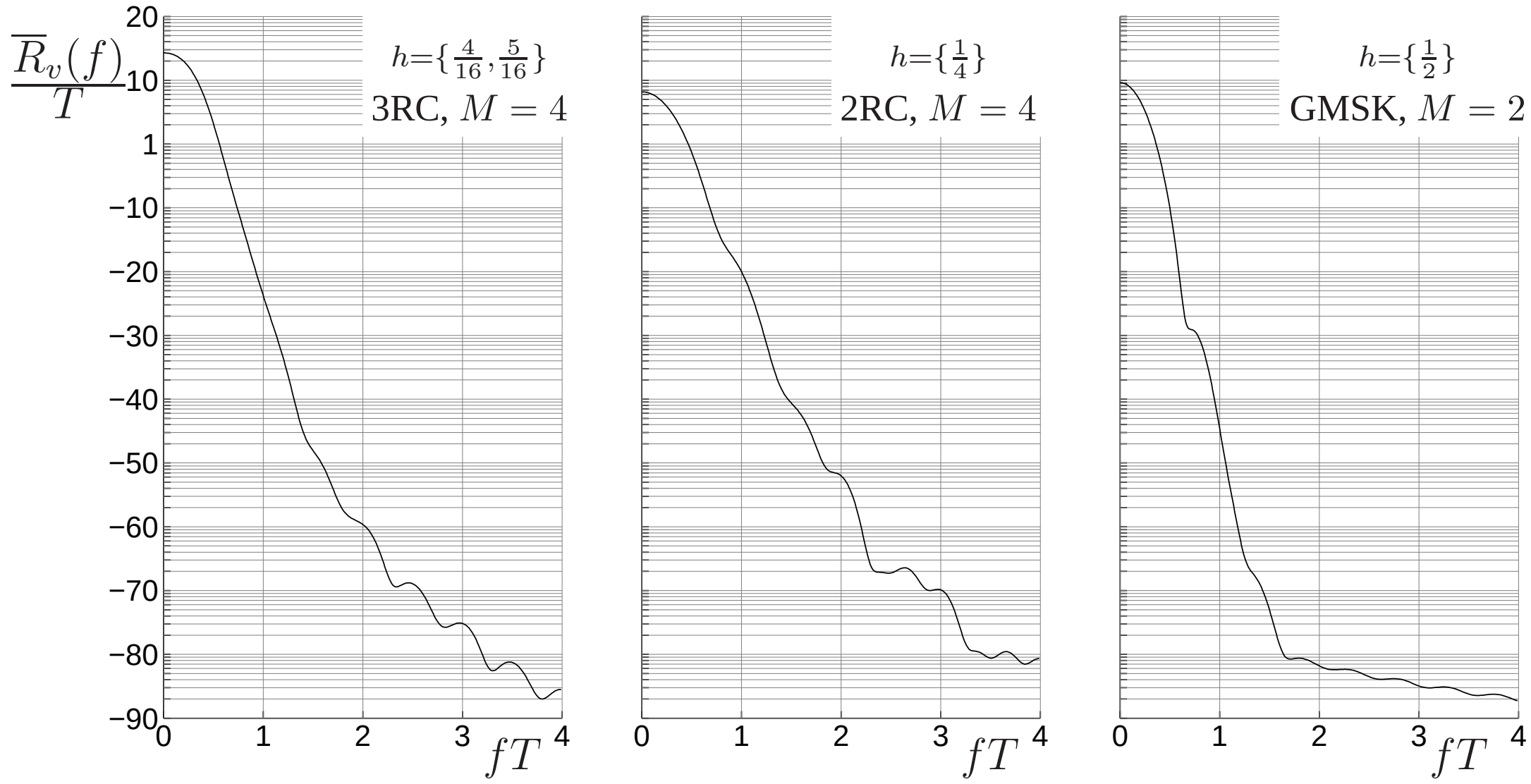

Fig. 10 – Normalised PSD plots for partial response single-$h$ and multi-$h$ CPM signals (equally likely input symbols).

# Appendix

## A. Sketch of the proof of Proposition4

We proceed step by step. For $L = 1$ we have

$$\left[\delta_{i_0,(j_0+r_n\alpha)_{2p}}\right]_{i_0,j_0\in\mathcal{N}_{2p}} = \boldsymbol{D}_{2p}^{r_n\alpha}$$

which is a square matrix obtained by cyclically shifting the main diagonal to the left by $r_n\alpha$ positions. The result can be obtained by evaluating the $r_n\alpha$ power of the single step ciclical shift matrix $\boldsymbol{D}_{2p} = \boldsymbol{D}_{2p}^1$. This corresponds to (34).

For $L = 2$, we have

$$\left[\delta_{i_0,(j_0+r_{n-1}j_1)_{2p}}\,\delta_{i_1,\alpha}\right]_{i_0,j_0\in\mathcal{N}_{2p},\,i_1,j_1\in\mathcal{A}_M} = \left[\delta_{i_0,(j_0+r_{n-1}j_1)_{2p}}\right]_{i_0,j_0\in\mathcal{N}_{2p},\,j_1\in\mathcal{A}_M} \otimes \left[\delta_{i_1,\alpha}\right]_{i_1\in\mathcal{A}_M}$$

where the first term can be further explicited as

$$\begin{aligned}\left[\delta_{i_0,(j_0+r_{n-1}j_1)_{2p}}\right]_{i_0,j_0\in\mathcal{N}_{2p},\,j_1\in\mathcal{A}_M} &= \sum_{\beta\in\mathcal{A}_M}\left[\delta_{i_0,(j_0+r_{n-1}\beta)_{2p}}\,\delta_{\beta,j_1}\right]_{i_0,j_0\in\mathcal{N}_{2p},\,j_1\in\mathcal{A}_M}\\ &= \sum_{\beta\in\mathcal{A}_M}\left[\delta_{i_0,(j_0+r_{n-1}\beta)_{2p}}\right]_{i_0,j_0\in\mathcal{N}_{2p}} \otimes \left[\delta_{\beta,j_1}\right]_{j_1\in\mathcal{A}_M}\end{aligned}$$

to obtain

$$\boldsymbol{e}_{\alpha,n} = \sum_{\beta\in\mathcal{A}_M} \underbrace{\boldsymbol{D}_{2p}^{r_{n-1}\beta}}_{2p\times 2p} \otimes \underbrace{\boldsymbol{w}_\beta^T}_{1\times M} \otimes \underbrace{\boldsymbol{w}_\alpha}_{M\times 1}\ .$$

The prospect is similar for $L = 3$, where

$$\left[\delta_{i_0,(j_0+r_{n-2}j_1)_{2p}}\right]_{i_0,j_0\in\mathcal{N}_{2p},\,j_1\in\mathcal{A}_M} \otimes \left[\delta_{i_1,j_2}\right]_{i_1,j_2\in\mathcal{A}_M} \otimes \left[\delta_{i_2,\alpha}\right]_{i_2\in\mathcal{A}_M}$$

with the central matrix being an identity. The general result is thus

$$\boldsymbol{e}_{\alpha,n} = \sum_{\beta\in\mathcal{A}_M} \boldsymbol{D}_{2p}^{r_{n-L+1}\beta} \otimes \boldsymbol{w}_\beta^T \otimes \underbrace{\boldsymbol{I}_M \otimes \cdots \otimes \boldsymbol{I}_M}_{L-2} \otimes \boldsymbol{w}_\alpha$$

where the kronecker product of $L-2$ occurrences of $\boldsymbol{I}_M$ is simply $\boldsymbol{I}_{M^L}$. As a consequence, (32) is valid.

### B. Sketch of the proof of (36)

Being $\boldsymbol{\pi}_n$ expressed as a Kronecker product in (33), we look for an eigenvector with the Kronecker structure

$$\boldsymbol{p}_n = \underbrace{\boldsymbol{u}_0}_{2p\times 1} \otimes \underbrace{\boldsymbol{u}_1}_{M\times 1} \otimes \cdots \otimes \underbrace{\boldsymbol{u}_{L-1}}_{M\times 1}$$

providing the set of equations

$$\begin{cases} \left(\sum_{\beta\in\mathcal{A}_M} \boldsymbol{D}_{2p}^{r_{n-L+1}\beta} \otimes \boldsymbol{w}_\beta^T\right)(\boldsymbol{u}_0 \otimes \boldsymbol{u}_1) = \boldsymbol{u}_0 \\ \boldsymbol{I}_M \boldsymbol{u}_2 = \boldsymbol{u}_1 \\ \vdots \\ \boldsymbol{I}_M \boldsymbol{u}_{L-1} = \boldsymbol{u}_{L-2} \\ \boldsymbol{q} = \boldsymbol{u}_{L-1} \end{cases}$$

By solving the system we immediately obtain $\boldsymbol{u}_1 = \cdots = \boldsymbol{u}_{L-1} = \boldsymbol{q}$, plus the remaining equation

$$\sum_{\beta\in\mathcal{A}_M} \left(\boldsymbol{D}_{2p}^{r_{n-L+1}\beta}\boldsymbol{u}_0\right) \otimes \left(\boldsymbol{w}_\beta^T \boldsymbol{q}\right) = \sum_{\beta\in\mathcal{A}_M} \boldsymbol{D}_{2p}^{r_{n-L+1}\beta}\boldsymbol{u}_0\, q_\beta = \boldsymbol{u}_0$$

which is solved by $\boldsymbol{u}_0 = \frac{1}{2p}\mathbf{1}_{2p}$, where $\mathbf{1}_{2p}$ is a column vector of length $2p$ with all entries set to $1$. Note that the solution is valid independently on $r_{n-L+1}$, and provides (36).

### C. Validity of (62)

By exploiting the first of (45a) and (48), the matrix $\boldsymbol{Y}_{\boldsymbol{x}}$ is

$$\begin{aligned} \boldsymbol{Y}_{\boldsymbol{x}} &= \left[Y_{\boldsymbol{x}}(\boldsymbol{i},\boldsymbol{j})\right]_{\boldsymbol{i}\in\mathcal{A}^{N_c+L-1},\boldsymbol{j}\in\mathcal{S}} \\ &= \left[W_{2p}^{j_0}\,\delta_{i_0,j_1}\,\cdots\,\delta_{i_{L-2},j_{L-1}}\,\delta_{x_0,i_{L-1}}\,\cdots,\delta_{x_{N_c-1},i_{N_c+L-2}}\right]_{i_0,\ldots,i_{N_c+L-2}\in\mathcal{A}_M,j_0\in\mathcal{N}_{2p},j_1,\ldots,j_{L-1}\in\mathcal{A}_M} \end{aligned}$$

where $\boldsymbol{i} = [i_0,\ldots,i_{N_c+L-2}]^T$ and $\boldsymbol{j} = [j_0,\ldots,j_{L-1}]^T$. It is now straightforward to see that the equivalent Kronecker form is given by (62).

### D. Sketch of the proof of Proposition9

For $p > 1$, we exploit the Kronecker products in (55) and (63) in (65), and observe that $\boldsymbol{V}_p\mathbf{1}_p = 0$. We then immediately obtain $\boldsymbol{m}_{\boldsymbol{y}}(\pm) = \mathbf{0}$ thus assuring that *no* spectral lines are found, as stated in (66a).

Instead, for $p=1$ we have $\boldsymbol{V}_1\,\mathbf{1}_1 = 1\cdot 1 = 1$, in which case it is easy to show that

$$\boldsymbol{m}_{\boldsymbol{y}}(+) = \underbrace{\boldsymbol{q}\otimes\cdots\otimes\boldsymbol{q}}_{N_c+L-1}, \qquad \boldsymbol{m}_{\boldsymbol{y}}(-) = W_{2p}\,\boldsymbol{m}_{\boldsymbol{y}}(+)\ .$$

Then, from the equivalence

$$\begin{aligned}\boldsymbol{r}_{\boldsymbol{y}}{}^{(d)}(nT_s) &= \boldsymbol{m}_{\boldsymbol{y}}(+)\,\boldsymbol{m}_{\boldsymbol{y}}{}^{*}(+) = \boldsymbol{m}_{\boldsymbol{y}}(-)\,\boldsymbol{m}_{\boldsymbol{y}}{}^{*}(-)\\ &= \underbrace{[\boldsymbol{q}\,\boldsymbol{q}^{*}]\otimes\cdots\otimes[\boldsymbol{q}\,\boldsymbol{q}^{*}]}_{N_c+L-1}\end{aligned}$$

we straightforwardly obtain (66b), by recalling (58).

## REFERENCES

Anderson86

[1] J. B. Anderson, T. Aulin, and C. E. Sundberg, *Digital Phase Modulation*. New York: Plenum, 1986.

Vitetta06

[2] G.M. Vitetta and F. Pancaldi, “Equalization algorithms in the frequency domain for continuous phase modulations,”*IEEE Trans. Commun.*, Vol. 54, No. 4, pp. 648-658, April 2006.

Laurent86

[3] P.A. Laurent, “Exact and approximate construction of digital phase modulations by superposition of amplitude modulated pulses (AMP),” *IEEE Trans. Commun.*, Vol. COM-34, No. 2, pp. 150160, Feb. 1986.

Mengali95

[4] U. Mengali and M. Morelli, “Decomposition of $M$-ary CPM signals into PAM waveforms,” *IEEE Trans. Inf. Theory*, Vol. 41, No. 5, pp. 1265-1275, Sep. 1995.

Huang03

[5] X. Huang and Y. Li, “The PAM decomposition of CPM signals with integer modulation index,” *IEEE Trans. Commun.*, Vol. 51, No. 4, pp. 543-546, Apr. 2003.

Cariolaro05

[6] G. Cariolaro and A.M. Cipriano, “Minimal PAM decompositions of CPM signals with separable phase,” *IEEE Trans. Commun.*, Vol. 53, No. 12, p. 2011-2014, Dec. 2005.

Perrins06

[7] E. Perrins and M. Rice, “PAM decomposition of $M$-ary multi-$h$ CPM,” *IEEE Trans. Commun.*, to be published.

Anderson78

[8] J.B. Anderson and D.P. Taylor, “A bandwidth-efficient class of signal-space codes,” *IEEE Trans. on Inform. Theory*, Vol. IT-24, No. 11, November 1978.

Wilson81

[9] S.G. Wilson, R.C. Gaus, "Power spectra of multi-$h$ phase codes," *IEEE Trans. Commun.*, Vol. COM-29, No. 3, pp. 250-256, March 1981.

Mazur81

[10] B.A. Mazur and D.P. Taylor, "Demodulation and carrier synchronization of multi-$h$ phase codes," *IEEE Trans. on Comm.*, Vol. COM-39, No. 3, March 1981.

Allen03

[11] J.C. Allen and B.E. Wahlen, "Multi-$h$ CPM Synchronization in Military Channels. Phase 2: A Simulation Framework," *Technical Report 1909*, SPAWAR System Center, San Diego, Sept. 2003.

Perrins05

[12] E. Perrins and M. Rice, "A new performance bound for PAM-based CPM detectors," *IEEE Trans. Commun.*, Vol. 53, No. 10, pp. 1688-1696, Oct. 2005.

Ilyin04

[13] D. Ilyin, "Spectra calculation of CPM signals," *Proc. of the IEEE Int. Conf. on Modern Problems of Radio Eng., Telecom. and Computer Science 2004*, pp. 195-197, 24-28 Feb. 2004.

Cariolaro06

[14] G. Cariolaro and A. Vigato, "Representation of a CPM modulator through a finite-state sequential machine," *submitted to ???*.

CarTron74

[15] G.L. Cariolaro, G.P. Tronca, "Spectra of block coded signals," *IEEE Trans. on Commun.*, Vol. 22, No. 10, pp. 1555-1564, Oct. 1974.

Yang98

[16] Y. Yang "General method for computing the power spectrum density of irreducible periodic Markov chains," *IEEE GLOBECOM 98*, Vol. 6, pp. 3320-3325, 8-12 Nov. 1998.

Gantmacher

[17] F.R. Gantmacher, *The theory of matrices*. New York: Chelsea Publishing Company, 1977.

Comm81

[18] "Special selection on combined modulation and encoding," *IEEE Trans. on Comm*, Vol. COM-29, No. 3, March 1981.

Proakis95

[19] J.G. Proakis, *Digital communications*, 3rd edition. Singapore: Mc Graw Hill, 1995.

## CAPITOLO 0 (sp) (June 28, 2018 11:26:07 )

---

File [ sp]

---

File [sp1]

---

File [bsp]